# Non-equilibrium, weak field induced cyclotron motion: a mechanism for magnetobiology

Ashot Matevosyan [1,2], Armen E. Allahverdyan [2,3]

[1] University of Cambridge, Cavendish Laboratory, 19 J.J. Thompson avenue, Cambridge CB3 0HE, UK

[2] Alikhanyan National Laboratory (Yerevan Physics Institute), Yerevan, Armenia.

[3] Yerevan State University, Yerevan, Armenia.

**Abstract**. There is a long-time quest for understanding physical mechanisms of weak magnetic field interaction with biological matter. Two factors impeded the development of such mechanisms: first, a high (room) temperature of a cellular environment, where a weak, static magnetic field induces a (classically) zero equilibrium response. Second, the friction in the cellular environment is large, preventing a weak field to alter non-equilibrium processes such as a free diffusion of charges. Here we study a class of non-equilibrium steady states of a cellular ion in a confining potential, where the response to a (weak, homogeneous, static) magnetic field survives strong friction and thermal fluctuations. The magnetic field induces a rotational motion of the ion that proceeds with the cyclotron frequency. Such non-equilibrium states are generated by a white noise acting on the ion additionally to the non-local (memory-containing) friction and noise generated by an equilibrium thermal bath. The intensity of this white noise can be weak, i.e. much smaller than the thermal noise intensity.

# 1. Introduction

The influence of a weak, static magnetic field on biological systems remains a controversial subject [1, 2, 3, 4]. A weak diamagnetism is present in such systems [3], but it can produce visible effects only for high magnetic fields (∼ 20 T) [5]. Experimental reports on the existence of weak-field biological influences are too numerous to be wrong [1, 2, 4]. However, they are frequently not reproducible, which asks for an explanation [1, 2, 4]. Physical mechanisms that would describe such influences in a sufficiently general molecular biology situation are unclear [6, 7, 8, 9].

Here we plan to study the influence of a weak static magnetic field on the stochastic motion of an ion (Brownian charge). Metal ions ($Na^+$, $K^+$, $Ca^{2+}$, $Mg^{2+}$ *etc*) play an important role in molecular biology. They can be called its third ingredient along with DNA/RNA and proteins [10]. Nearly 1/3 of all proteins employ metal ions for their functioning [3, 10]. Ions are important in bioenergetics, communication (e.g., nerve impulse generation), osmotic regulation, metabolism, energy storage *etc* [3, 10, 11, 12, 13]. Hence, it is natural to study their motion as a target of a static magnetic field. Such a study should look at non-equilibrium states, since under weak magnetic fields ion's translational motion is *classical*, and then an *equilibrium* magnetic response (moment) is nullified by the Bohr-van Leeuwen theorem [14, 15, 16, 17, 18, 19, 20]. Its message is straightforward: once the magnetic field does not do work on the charge, it does not influence charge's equilibrium state that depends only on its energy.

The main issue of envisaging possible non-equilibrium responses of a static magnetic field is that there is a huge difference between the relaxation time induced by friction (<$10^{-9}$ s for ions in water) and the magnetic (cyclotron) time-scale (>$10^{-3}$ s for weak fields) [6, 8]. For example, consider the free ion diffusion (i.e., Brownian motion without an external potential) that is a pertinent non-stationary (hence non-equilibrium) cellular process [3]. Theoretically, it is influenced by a static magnetic field, since the diffusion is impeded in directions perpendicular to the field due to a circular (cyclotron) motion induced by the magnetic field [21]. But for weak fields this effect is completely diminished by large translation friction (damping) in water: the influence of the magnetic field on the perpendicular diffusion is or order of $b^2/\gamma^2$, where $b$ and $\gamma$ are (resp.) the cyclotron frequency induced by the magnetic field and the friction coefficient in water. For the Earth magnetic field we get $b^2/\gamma^2 \sim 10^{-18}$ ; see (2.8).

Models that attempt to explain the effects of weak magnetic fields postulate the existence of degrees of freedom moving under a weak (or without) friction [2, 22, 23, 24, 25, 26]. Such postulates are so far unfounded, since the friction is large both in water and inside proteins cavities, and its differences (e.g. due to hydrophobic effects) are too small to account for the above big gap [6, 8, 9].

We aim to propose mechanisms for the influence of weak static magnetic fields that survives strong friction and room temperature fluctuations. To this end, we study an ion confined in a potential generated by a protein or membrane. Cellular ions are bound to proteins or diffuse freely in the cell; we focus on the former type of ion motion [3, 11, 10]. Ion's interaction with the thermal environment generates friction and noise, in addition to the regular force exerted by the confining potential [3, 27, 28]. The potential is strongly confining for an ion in a tight cavity, while it is weaker if the ion is bound next to protein surface. If these three forces—together with the Lorentz force generated by the magnetic field—were the only ones acting on the ion, then its motion will quickly reach an equilibrium state that does not feel the static magnetic field (the Bohr-van Leeuwen theorem); in particular, no cyclotron motion will be present. However, we take into account that the chaotic cellular environment is capable of generating additional noises, e.g. via fast conformational motion of protein [29, 30, 31, 32, 33, 34]. We make natural assumptions that the additional noise is white, i.e., it is generated by fast degrees of freedom, and that it is weak: the white-noise intensity is much smaller than that of the thermal noise. Accounting also for the memory necessarily present in the aqueous thermal bath, we find a stationary state, where the ion rotates with the cyclotron frequency in a weak field and an aqueous environment. The origin of this effect is that although the additional white noise leads to a stationary state that is quantitatively close to the equilibrium state in many of its features, it is non-equilibrium and involves violations of Fluctuation-Dissipation Relation (FDR). Due to this, the Bohr-van Leeuwen theorem does not generally hold out of equilibrium [18, 19]. The existence of FDR violations is well-established in active biophysical processes; see [35, 36] for reviews.

Thus, our model explains how the average cyclotron motion can be sustained in the frictional and noisy cellular environment. Note that many functional processes in cation-driven proteins proceed on times comparable with the cyclotron motion in weak magnetic fields [31]. Hence, the cyclotron motion can influence such processes, e.g. via imposing the cylindrical symmetry of the cyclotron motion on the shape of the ion-biding protein cavity [24].

We aimed to organize this paper such that its main result is available to both experimentalists and theoreticians. Hence, we relegated detailed derivations of our results to Appendices making our conclusions self-contained. The next section discusses our basic model: an ion interacting with an equilibrium thermal bath and subjected to an external white noise. Section 3 shows that the magnetic field induces a cyclotron motion and diamagnetic response to a static magnetic field. Section 4 studies stochastic trajectories numerically for several ranges of parameters, explains them via autocorrelation functions, and confirms our analytical results. Here we study overdamped and underdamped regimes of the stochastic dynamics and show that the memory present in the aqueous thermal bath facilitates underdamping. We summarize in the last section with discussing again the main assumptions of the model and providing a perspective on the future research. This section also reviews two other physically consistent approaches for explaining influences of weak, static magnetic fields, viz. the compass mechanism and radical pair reactions. Appendix A compares the obtained non-equilibrium classical magnetic moments with known scenarios of equilibrium and dissipative quantum magnetism. Appendix B discusses relations between Langevin and Fokker-Planck equations, while Appendices C—D describes our technical tools and outlines generalizations of our results.

## 2. The model

### 2.1 Langevin equation

We study ions bound in external potentials (generated by membranes or proteins). Hence we shall neglect electrostatic interaction between ions [3], focusing instead on their interaction with thermal bath and with external forces. Thus we describe a single ion with mass $m$, charge $Q$, coordinate vector $\mathbf{x}(t) = (x, y, z)$, and velocity $\mathbf{v} = \dot{\mathbf{x}} \equiv \frac{d\mathbf{x}}{dt}$. The Langevin equation divided over the mass $m$ reads [27, 28, 37]:

$$\dot{\mathbf{v}}(t) = -\gamma \int_{t_0}^{t} dt' \, \kappa \, e^{-\kappa|t-t'|} \, \mathbf{v}(t') - u_\mathbf{x}(\mathbf{x}) + \frac{Q}{m} \mathbf{v} \times \mathbf{B} + \frac{1}{m} \boldsymbol{\eta}(t) + \frac{1}{m} \boldsymbol{\xi}(t), \qquad (2.1)$$

where $Q \, \mathbf{v} \times \mathbf{B}$ is Lorentz's force coming from an external static magnetic field $\mathbf{B}$ (here $\times$ stands for the vector product), $u_\mathbf{x}(\mathbf{x}) \equiv \partial u/\partial \mathbf{x}$ is the force generated by a potential $u(\mathbf{x})$, and where the friction force with magnitude $\gamma$ and exponential kernel has a well-defined memory time $1/\kappa$. In (2.1) $\boldsymbol{\eta}(t)$ and $\boldsymbol{\xi}(t)$ are independent Gaussian noises, where $\boldsymbol{\eta}(t)$ is generated by the equilibrium thermal bath, while $\boldsymbol{\xi}(t)$ is generated by a (random) time-dependent potential $-(x\xi_x + y\xi_y + z\xi_z)$, which operates in addition to the deterministic potential $u(\mathbf{x})$ and refers e.g. to fast conformational motions of the protein; see below for details.

The initial conditions for (2.1) are posed at $t = t_0$. Parameters $\gamma$, $\kappa$ and $Q|\mathbf{B}|/m$ in (2.1) have dimension of frequency. Here $Q|\mathbf{B}|/m$ is the well-known cyclotron frequency [2, 4], and $m\gamma$ is the proper friction constant. Another frequency comes via $u(\mathbf{x})$, e.g., the isotropic harmonic potential brings frequency $\omega_0$:

$$u(\mathbf{x}) = \frac{\omega_0^2}{2}|\mathbf{x}|^2 = \frac{\omega_0^2}{2}(x^2 + y^2 + z^2), \tag{2.2}$$

where we note that the $\mathbf{x}$ coordinate system is located at the bottom of the potential. Let $\delta(t)$ and $\delta_{ij}$ be Dirac's delta-function and Kronecker's symbol, respectively. The averages $\langle \dots \rangle$ of the Gaussian noises read [27, 28]:

$$\langle \eta_i(t) \rangle = 0, \qquad \langle \xi_i(t) \rangle = 0, \qquad i = x, y, z, \tag{2.3}$$

$$\langle \eta_i(t)\eta_j(t') \rangle = \delta_{ij} q \ (\theta/2) \ e^{-\theta|t-t'|}, \qquad \langle \xi_i(t)\xi_j(t') \rangle = q_w \delta_{ij} \delta(t - t'). \tag{2.4}$$

Here $q$ and $q_w$ are noise intensities, $1/\theta$ is the correlation time of the $\boldsymbol{\eta}$-noise, while the $\boldsymbol{\xi}$-noise is white. Recall that if in (2.1) the friction ($\propto \gamma$) and the noise $\boldsymbol{\eta}(t)$ are generated by the same *equilibrium* thermal bath, then the fluctuation-dissipation relation (FDR) holds [28, 37], i.e., the friction memory $e^{-\kappa|t-t'|}$ and the noise correlation $e^{-\theta|t-t'|}$ are the same:

$$\theta = \kappa. \tag{2.5}$$

Another aspect of the FDR is that the intensity of the $\boldsymbol{\eta}(t)$-noise relates to the thermal bath temperature $T$ ($k_B$ is Boltzmann's constant) [27, 28]:

$$q = 2m\gamma k_B T. \tag{2.6}$$

We shall assume that (2.5, 2.6) hold, and then the only non-equilibrium aspect in the problem comes from the external white noise $\boldsymbol{\xi}(t)$ that does not satisfy FDR, since it is not accompanied by the corresponding friction (even if there would be the additional friction, the FDR can be still violated due to different temperatures of the noises).

Equations similar to (2.1), i.e., Langevin equations with magnetic field, were studied at many places in application to plasma physics or stochastic systems [18, 19, 20, 21, 38, 39, 40, 41, 42, 43, 44]. Our focus will be on implications of (2.1) for ions moving in cellular environment having a large friction ($\gamma \sim 10^{12}$ s$^{-1}$) and thermal fluctuations ($T = 300$ K).

### 2.2 Friction, memory, noise and potential: Magnitudes of involved parameters

For the Brownian ion moving in a viscous fluid (water) the friction (i.e., energy dissipation) is generated by fluid's shear viscosity [37, 45, 46]. For the ion with mass $m$ the friction constant $m\gamma$ is estimated via the Stokes-Einstein formula [27, 28, 45, 46]:

$$m\gamma \sim r_i \eta, \qquad (2.7)$$

where $r_i$ is the (effective) radius of the ion, and $\eta$ is the shear viscosity of the fluid; $\eta = 10^{-2}$ g cm/s for water at room temperature. Important biological ions (Na$^+$, K$^+$, Ca$^{2+}$, Mg$^{2+}$) have roughly the same mass and size; e.g., for Ca$^{2+}$ we have $m = 40$ g/mol $= 6.7 \times 10^{-23}$ g, $r_i \sim 0.2$ nm and we get $\gamma \sim 10^{13}$ s$^{-1}$ from (2.7). Ions bound inside of protein interact with both protein degrees of freedom and water. The protein viscosity is close to that of water, and we find the same result $\gamma \sim 10^{12} - 10^{13}$ s$^{-1}$ [47]. A similar estimate for $1/\gamma$ is obtained by assuming that the ion is bound inside of a cage of linear size $l$, and interacts with thermal motion of cage's walls. Then $1/\gamma \sim l/v_{\text{th}} = l/\sqrt{k_B T/m}$, which for $l = 0.5$ nm produces $\gamma \sim 10^{12}$ s$^{-1}$ [6].

The standard division of static magnetic field magnitudes is as follows [1, 4]: weak ($\leq 1$ mT), moderate ($> 1$ mT and $\leq 1$ T), high ($> 1$ T and $\leq 20$ T), ultra-high ($> 20$ T). Modern NMR medicine operates in the strong field range $1 - 3$ T, while the Earth magnetic field is $\sim 50\mu$T. For definiteness, we shall assume $B = 1$ mT, which for Ca$^{2+}$ with $m = 6.7 \times 10^{-23}$ g, and the charge $Q = 3.2 \times 10^{-19}$ Cl leads to the cyclotron frequency $b = \frac{QB}{m} \sim 5 \times 10^3$ s$^{-1}$. The choice of $B = 1$ mT is motivated by its experimental relevance [1, 4], and by the fact that is the strongest among weak fields. This choise is 20 times larger than the Earth magnetic fields, which is convenient experimentally since the latter is normally always present [4].

Equation (2.1) contains memory described by the simplest kernel $\kappa e^{-\kappa|t-t'|}$ with a well-defined memory time $1/\kappa$. The memory-less situation is recovered for $\kappa \to \infty$, where we get $\kappa e^{-\kappa|t-t'|} \to 2\delta(t-t')$ in (2.1), and hence the memory-less friction $-\gamma \mathbf{v}$ [28, 37]. The general mechanism for the memory is that environmental relaxation processes have a finite characteristic time, which (roughly) agrees with the memory time $1/\kappa$ [45, 46]. Hence, for a Brownian ion moving in water, the friction memory is present, since water molecules and the ion have comparable masses [48, 49]. The specific form $e^{-\kappa|t-t'|}$ in (2.1) holds for a particle diffusing in a visco-elastic medium [50]. Such media describe the cell interior [51, 52]. A virtue of the kernel $e^{-\kappa|t-t'|}$ is that we can construct from (2.1) a Fokker-Planck equation also for any (non-linear) potential; see Appendix B. Note that the hydrodynamic theory of the friction memory leads to kernels that are more complex than $e^{-\kappa|t-t'|}$ in (2.1). This theory is however not yet complete [53].

For an aqueous environment the memory time $1/\kappa$ can be estimated via the cage-jump model [54, 55], where a given molecule diffuses by jumping from one cage (made by neighbouring molecules) to another. The characteristic time a molecule spends in one cage is the memory time. This time can be estimated as $\tau_0 e^{\Delta/(k_B T)}$, where $\Delta$ is the energy barrier (including the hydrogen bonding contribution) that defines the cage, $\tau_0 \sim 10^{-13}$ s, and $\Delta = 0.2$ eV [55]. This estimate for $\tau_0$ can be obtained via the linear size $l_c \simeq 0.1$ nm of the cage and thermal velocity $v_{\text{th}}$ of cage's walls: $\tau_0 \sim l_c/v_{\text{th}} = l_c/\sqrt{k_B T/m_w}$, where $m_w$ is the mass of a water molecule. Recalling $k_B T = 0.025$ eV at room temperature, we get for the memory time $\kappa^{-1} \sim \tau_0 e^{\Delta/(k_B T)} \sim 10^{-10}$ s [3]. This is in between of the hydrogen bond rearranging time $10^{-11}$ s and the water structural reorganization time $10^{-8}$ s [56].

Once the memory in friction is understood, the structure of the $\boldsymbol{\eta}$-noise in (2.51) is recovered through the FDR (2.5). The $\boldsymbol{\xi}$-noise in (2.1) is generated by fast conformational motions of protein or membrane that interact with the bound ion [29, 30]. These motions are described as noise [57]. We assume that this noise is white, since the corresponding degree of freedom have small characteristic times ($10^{-15}$ - $10^{-13}$ s) [31]. Recall that $\boldsymbol{\xi}$ does not hold FDR.

For estimating the frequency $\omega_0$ of the harmonic potential $m\omega_0^2\mathbf{x}^2/2$ in (2.2), we note that that since the confined ion is still subject to thermal fluctuations, we can take $m\omega_0^2 l^2/2 \sim k_\mathrm{B} T$, where $l$ is the characteristic localization length. The confining potential (2.2) can describe two different scenarios. The first scenario is when the ion is localized next to the protein (or membrane) surface. For this case we can estimate $l \sim 1$—$1.5$ nm. The second scenario is when the ion is tightly bound inside of a protein cavity. Then $l$ is a few times larger than the ion radius: $l \sim 0.5$ nm, hence $\omega_0 \sim 10^{11}$ s$^{-1}$. Note that for the harmonic potential (2.2) the probability density of the coordinate $\mathbf{x}$ (in the stationary situation) is Gaussian; see Appendix D. Hence, we can take $l = \sqrt{\langle \mathbf{x}^2 \rangle - \langle \mathbf{x} \rangle^2}$, since the particle is nearly absent for distances $|\mathbf{x}| > l$ (note that $\langle \mathbf{x} \rangle = 0$).

Since we are interested in weak magnetic fields, we use in our estimates $B \sim 1$ mT, and the room temperature 300 K of the biological environment. Thus, we accept for further estimates the following hierarchy of inverse times ($m = 6.7 \times 10^{-26}$ kg, and the charge $Q = 1.6 \times 10^{-19}$ Cl):

$$\gamma \sim 10^{12} > \omega_0 \sim 10^{11} > \kappa \sim 10^{10} \gg b = QB/m \sim 10^3 \text{ s}^{-1}. \tag{2.8}$$

Finally, we should estimate the white noise intensity $q_w$ in (2.4) (the thermal nose intensity $q$ is fixed from (2.6)). We should require that $q_w$ is sufficiently small compared to $q$, otherwise the thermal averages (e.g. $\langle \mathbf{x}^2 \rangle$) will change significantly. Given (2.8) we can assume $q_w$ as small as $q_w \simeq 10^{-2} q$. We cannot take $q_w$ much smaller than this value, since then the influence of the white noise will be negligible. Note that the assumption on the white feature of $\boldsymbol{\xi}$ is relaxed in Appendix C.

## 3. Diamagnetism in harmonic potential

Langevin equation (2.1) can be converted into Fokker-Planck equation for the joint probability of the involved random variables; see Appendix B. We look for the stationary regime, where the probability density of random variables pertaining to the ion motion becomes independent from time [28, 37]. For a particle in a confining (e.g. harmonic) potential this stationary regime is reached from any state after a short relaxation time that scales as $10^{-9}$ s for parameters given by (2.8); see (4.3). The solution of the stationary Fokker-Planck equation is described in Appendix C. Here we shall discuss this solution for the simplest case of the harmonic potential (2.2). More general potential can (and must) be studied, but the harmonic potential suffices for introducing and discussing our main result.

We choose the magnetic field along the z-axis in (2.1): $\mathbf{B} = |\mathbf{B}|\mathbf{e}_z$. In the stationary state all linear averages nullify: $\langle x \rangle = \langle y \rangle = 0$ (due to (2.2)) and $\langle v_x \rangle = \langle v_y \rangle = 0$, because the motion is confined. We get for the averages of the orbital momentum (along the magnetic field) $\langle L \rangle$, square coordinate $\langle x^2 + y^2 \rangle$ in the plane perpendicular to the magnetic field, and $\langle v_x^2 + v_y^2 \rangle$ (see Appendix C):

$$\langle L \rangle = \langle xv_y - yv_x \rangle = -b\frac{q_w}{\gamma m^2 \kappa^2}, \tag{3.1}$$

$$\langle v_x^2 + v_y^2 \rangle = \frac{q}{\gamma m^2} + \frac{q_w}{\gamma m^2} \times \frac{\kappa^2 + \omega_0^2 + \gamma\kappa + b^2}{\kappa^2}, \tag{3.2}$$

$$\langle x^2 + y^2 \rangle = \frac{q}{\gamma m^2 \omega_0^2} + \frac{q_w}{\gamma m^2 \omega_0^2} \times \frac{\kappa^2 + \omega_0^2}{\kappa^2}, \tag{3.3}$$

where $b$ denotes the cyclotron frequency:

$$b = QB/m. \tag{3.4}$$

Equations (3.1—3.3) are special cases of the harmonic potential solution derived in Appendix C; see (C.26). However, (3.1) follows directly from the more general argument discussed after (C.24) that applies to anharmonic potentials as well.

Note that the averages in (3.1-3.3) hold symmetry features, e.g., $\langle xv_y \rangle = -\langle yv_x \rangle$, $\langle v_x^2 \rangle = \langle v_y^2 \rangle$ etc; see Appendix C. Equation (3.1) shows that the response to the magnetic field is diamagnetic, i.e., $\frac{\langle L \rangle}{B} < 0$. For the considered harmonic potential (2.2), $\langle L \rangle$ does not depend on the shape $\omega_0^2$ of the potential and on the intensity $q$ of the thermal noise; cf. (3.1). Such dependences emerges for anharmonic potentials [see (C.21) in Appendix C], but the fact of diamagnetic response stays.

Now $\langle L \rangle$ nullifies for three cases; see (3.1). First, $\langle L \rangle = 0$ when $\kappa \to \infty$ in (3.1), since now we get that for long times the system described by (2.1) holds FDR with white noises and memory-less friction (with a certain effective temperature). Hence it relaxes to an equilibrium state that does not contain the magnetic field $b$ (the Bohr-van Leeuwen theorem), and leads to $\langle L \rangle = 0$. In the same limit $\kappa \to \infty$ we get from (C.23) the known thermal expression for $\langle v_x^2 + v_y^2 \rangle = 2\langle v_x^2 \rangle$. Second, $\langle L \rangle = 0$ for $q_w = 0$ (zero intensity of the white noise) where FDR again holds. Third, $\langle L \rangle = 0$ for $b \to 0$ (zero magnetic field). Note that $\langle L \rangle$ increases in a (unrealistic) limit $\kappa \to 0$, because now the friction term nullifies, as seen from (2.1). For a nullifying friction the finite-intensity white noise brings a large amount of energy into the system, forcing the oscillator to move faster over trajectories more remote from the origin, and thereby increasing $\langle v_x^2 + v_y^2 \rangle$, $\langle x^2 + y^2 \rangle$ and also $\langle L \rangle$; see (3.1-3.3).

The average orbital momentum in (3.1) can be expressed via magnetic moment $\mathcal{M}_c = Q\langle L \rangle/2$ [16], where $Q$ is ion's charge. Using (2.6) we find for the magnetic moment:

$$\mathcal{M}_c = -\frac{q_w}{q} \frac{k_B T Q^2 B}{m^2 \kappa^2}. \tag{3.5}$$

Since $\langle x^2 + y^2 \rangle$ is the square of the characteristic radius, the (average) angular velocity $\langle \Omega \rangle$ that corresponds to $\langle L \rangle$ in (3.1) is obtained as

$$\langle \Omega \rangle = \frac{\langle xv_y - yv_x \rangle}{\langle x^2 + y^2 \rangle} = -b \left[ \left(1 + \frac{q}{q_w}\right) \frac{\kappa^2}{\omega_0^2} + 1 \right]^{-1}. \tag{3.6}$$

We confirm in Appendix D that (3.6) is indeed the average of the angular velocity

$$\Omega = \frac{xv_y - yv_x}{x^2 + y^2}. \tag{3.7}$$

It is seen from (3.6) that $\langle \Omega \rangle/(-b)<1$. Using (2.8) we conclude that even for $\frac{q_w}{q}$ as small as $\sim 10^{-2}$ (see the discussion after (2.8)) we can get $\langle \Omega \rangle/(-b) \sim 1$, because $\frac{\kappa^2}{\omega_0^2} \sim 10^{-2}$ in (3.6). Thus the average angular velocity $\langle \Omega \rangle$ is of the same order of magnitude as the cyclotron frequency $b$, despite the large friction and the strong thermal noise (stronger than the non-equilibrium noise).

The time-scale $1/b$ of the cyclotron motion is much larger than the other characteristic times including the relaxation time; see (2.8) and the discussion after it. Hence (3.6) describes a rotation of a probability cloud, rather than a rotation along a well-defined orbit; see section 5 for more details on this. The linear velocity of the cyclotron motion is estimated as $1 \text{ nm} \times 10^3 \text{ s}^{-1} \sim 10^{-6}$m/s (see (2.8)) that by its order of magnitude coincides with the ordered motion velocity inside of the cell [28, p 24]. In contrast to (free) diffusion, such an ordered motion is driven by active sources via energy dissipation [28]. Times $\sim 10^{-3}$ s are typical for the functional activity of many metalloproteins [31], e.g., for $Ca^{2+}$ specific binding to calmodulin [58]. Hence (3.6) supports the hypothesis that the additional ordered motion induced by a weak magnetic field can influence functional activity of proteins [24]. In particular, the influence can be realized via imposing on cation-binding cavity (calmodulin) the cylindrical symmetry that is inherent in the average cyclotron motion. The estimates for the angular speed given in [24] do not support the hypothesis, since they do not account for the strong friction and thermal noise. Note that if several cations are bound in a protein (e.g., calmodulin can bound up to four $Ca^{2+}$), then the collective average orbital moment of all the cations equals the sum of individual contributions given by (3.1).

We emphasize that the application of (3.1) to magnetobiological effects *is not* that it corresponds to a directly observable collective magnetic moment generated via (3.5); i.e. the application is based on $\langle \Omega \rangle$ in (3.6), and not on the magnetic moment $\mathcal{M}_c$ in (3.5). Appendix A compares (3.5) with the quantum diamagnetic moment that is produced by electron motion inside of atoms, a quantum effect responsible for the water diamagnetism [15]. We conclude that the magnitude of (3.5) is smaller than the electronic diamagnetic moment of water molecule. The physical content of these diamagnetic effects is quite different: (3.1) produces an ordered motion with a velocity reasonable for a strongly frictional cell environment, while the quantum diamagnetism relates to a zero-temperature intra-atom motion of frictionless electrons. At any rate, both (3.5) and the water diamagnetism are very small for the considered range of small magnetic fields. Appendix A also compares (3.5) with the spin paramagnetic moment at the same temperature $T$, and concludes that (3.5) can be larger than this paramagnetic moment.

## 4. Autocorrelation, charateristic times, and numerical results

### 4.1 Relaxation and rotation times

Above we calculated features of the stationary state for the ion. Now we turn to studying the stochastic ion's motion from the viewpoint of random trajectories. We focus on $x$ and $y$ coordinates of (2.1) under harmonic potential (2.2) and introduce the complex coordinate $\zeta = x + iy$. The advantage of using $\zeta$ is that rotations on the $xy$ plane are described via the phase of $\zeta$.

The autocorrelation function $\langle \zeta(t_1)\zeta^*(t_2) \rangle$ studies the stochastic motion in the time-window $(t_1, t_2)$, assuming that $t_1$ and $t_2$ are much larger than the relaxation time, but $t_1 - t_2$ can be arbitrary. The behavior of $\langle \zeta(t_1)\zeta^*(t_2) \rangle$ depends on the involved time-scales, and allows to distinguish between underdamping, where on times smaller than the relaxation times the ion makes rotations, from overdamping (no rotations). We deduce (see Appendix E for the derivation):

$$\langle \zeta(t_1)\zeta^*(t_2)\rangle = \sum_{n=1}^{3} C_n e^{i\omega_n t} = \sum_{n=1}^{3} C_n e^{-u'_n t - i u''_n t}, \qquad t = |t_1 - t_2|, \qquad (4.1)$$

$$u_n = u'_n + i u''_n, \qquad n = 1,2,3, \qquad (4.2)$$

where $u_n$ (with real part $u'_n$ and imaginary part $i u''_n$) are the roots of the characteristic equation:

$$u^3 - (\kappa + ib) u^2 + u(\gamma\kappa + \omega_0^2 + ib\kappa) - \omega_0^2\kappa = 0. \qquad (4.3)$$

$C_n$ in (4.1) are $t$-independent pre-factors that are determined from the initial conditions at $t_1 = t_2$; cf. (3.1). Autocorrelation functions of other quantities are expressed as in (4.1), but with different pre-factors $C_n$. We are not interested in $C_n$. We emphasize that (4.3) does not depend on the noises $\boldsymbol{\eta}(t)$ and $\boldsymbol{\xi}(t)$ in (2.1).

Thus $u'_n > 0$ in (4.1, 4.2) correspond to inverse relaxation times, while $u''_n$ refer to rotation frequencies. Indeed, for $t \gg u'^{-1}_n$, we get $\langle \zeta(t_1)\zeta^*(t_2)\rangle = \langle \zeta(t_1)\rangle\langle \zeta^*(t_2)\rangle = 0$. We also refer to $u''^{-1}_n$ as rotational period, leaving out $2\pi$ factor. We have rotational (underdamped) motion if $u'_n \ll u''_n$ holds for at least one index $n$. Otherwise, the motion is overdamped. Consider a particular simple case in (4.3). For $\kappa \gg 1$, i.e., in the memoryless friction regime, (4.3) leads to $u'_1 \simeq \kappa \gg u''_1, u'_2, u''_2, u'_3, u''_3$. Hence the the contribution from $n = 1$ quickly disappears from (4.1), and we are left with two roots of the quadratic equation $u^2 - u(\gamma + ib) - \omega_0^2 = 0$ found from (4.3). For a small $b$, these two roots predict that the transition from overdamping to underdamping takes place at $\gamma \sim 2\omega_0$. For not so large $\kappa$ the underdamping is facilitated, since the friction $\gamma$ in (4.3) is multiplied by $\kappa$. Note that when parameters in (2.8) are put into (4.3) we get that relaxation times scale as $10^{-9}$ s.

## 4.1 Numerical results

We solved numerically the $(x, y)$ components of (2.1, 2.2), i.e., the planar motion under the magnetic field directed along the $z$ axis. The numerical solution employs the Euler-Maruyama method [28] with the time-step $10^{-4}$. The ratios of parameters are kept realistic, as in (2.8), except for the magnitude of the cyclotron frequency $b$ that is taken larger than its relative value in (2.8) for making numerically visible the influence of the magnetic field. However, $b$ is still 100 times smaller than other frequencies in (2.8). Hence, the reported numerical results allow to check our analytical results and also visualize the effect of a relatively weak magnetic field.

Figures 1–4 show realizations of random trajectories— both projected to the angle variable $\int_0^t \mathrm{d}s\, \Omega(s)$ given by (3.7) (first row of figures) and in the $(x, y)$-plane (second row)—for various ranges of parameters. Parameters of Figure 1 are those of underdamping due to memory: the relaxation times are larger than the rotation periods: $u''^{-1}_2 \ll u'^{-1}_2$ and $|u''_3|^{-1} \ll u'^{-1}_3$ in (4.1). For times smaller than the relaxation time(s), there are almost deterministic rotational trajectories with small stochastic effects. For times comparable with relaxation times, these trajectories start to shift randomly; see the second (bottom) row of Figure 1. The first (top) row of Figure 1 visualizes motion of the ion for longer time intervals, each color for a separate random trajectory. The vertical axis shows the angle rotated around the origin (0,0), and a constant slope means constant angular velocity $\Omega$. The slope is constant for times smaller than the relaxation time $u_3'^{-1} \simeq 2$. For some trajectories the slope can be constant over longer times. As shown in the first row of Figure 1, ions rotates randomly both clockwise and counterclockwise. However, the angular velocities are not the same and on average the ion will rotate clockwise due to the magnetic field, as predicted by (3.6). The empiric average —calculated via just 19200 random trajectories—agrees well with the theoretical average (3.6). Along a single trajectory the average cyclotron motion is seen as a slow rotation of the probability cloud generated by faster motions, since the cyclotron period is the longest time-scale.

Figure 2 shows a moderately damped situation, since $\omega_0$ is 9 times smaller than in Figure 1. Rotations are still visible. In contrast to Figure 1, the trajectories are more irregular; see the first row of Figure 2. Here the longest relaxation time $u'^{-1}_1$ corresponds to the component with the longest rotational period $u''^{-1}_1$ (very slow rotation). Indications of this are seen in the first row of Figure 2: some slopes are "flat" with zero rotation on average. Again, we reproduce the theoretical prediction (3.6) via averaging over random trajectories (in figures we employed 19200 random trajectories, though in certain cases a much smaller number of trajectories is sufficient). Figure 3 demonstrates a memory-less underdamped situation: now both $\kappa$ and $\omega$ are larger than $\gamma$, and certain relaxation times are larger than the rotation periods. Both Figure 3 and Figure 1 demonstrate rotational motion, but the underlying mechanisms are different: for Figure 3 it is due to a large $\omega_0$ (i.e., strongly confining potential), while for Figure 1 it is due to memory in friction. Note that trajectories in the second row of Figure 1 are more regular than those in Figure 3.

Figure 4 shows a memory-less situation (i.e., a large $\kappa$) and loosely confining potential (moderate $\omega_0$). In this overdamping regime, the relaxation times are much smaller than the rotation times $u''^{-1}_n \gg u'^{-1}_n$, $n=1,2,3$. This is the standard overdamping regime, because now (4.3) has 3 real roots under $b \to 0$: the large rotation periods are due to the small perturbation of the magnetic field. Hence no deterministic rotation is there on short times. The trajectories in the bottom row of Figure 4 resemble those of the ordinary random walk with a very slow convergence of the empiric $\int_0^t ds\, \Omega(s)$ to its theoretical value.

Note the spread of $\int_0^t ds\, \Omega(s)$ among Figures 1-4: it is the largest for Figure 1, where the short-time motion is rotational, and smallest for Figure 4, when there are no deterministic rotations and $\int_0^t ds\, \Omega(s)$ changes due to random walk.

## 5. Summary

### 5.1 Assumptions and Results

To understand the influence of a weak, static magnetic field on the stochastic motion of a cellular ion we adopted a Brownian (Langevin-equation based) model (2.1) with the following assumptions:

–The equilibrium thermal bath of the ion generates friction and noise. They relate to each other via the fluctuation-dissipation relation (2.5). Both friction and noise have a finite memory, which comes from the fact that the bath consists of particles with the size not much smaller than that of the ion (e.g., water molecules). Hence, the friction is not given by the ordinary (Ohmic) expression, and we accepted the simplest (exponential) model for the memory; see (2.1).

–The ion moves in a confining potential. This can be a tight protein cavity or a weaker confinement of the ion next to the protein surface.

–Besides the noise acting from the equilibrium bath, there is a weak white noise generated by non-equilibrium cellular processes, e.g., those coming from fast protein degrees of freedom; cf. (2.1).

Hence, the stationary state reached by the ion is not equilibrium, though in absence of the magnetic field it does have equilibrium features, since the non-equilibrium (white) noise is weak.

In this model we found the following results:

• A weak static magnetic field induces a non-zero average orbital moment (3.1) (diamagnetic response to the magnetic field), and leads to rotation with cyclotron frequency (3.6). The cyclotron period has the same order of magnitude (millisecond) as functional motion of many proteins [31]. Likewise, the linear velocity that corresponds to this cyclotron motion has the same order of magnitude as other ordered motions in the cell. These results support the hypothesis by [24] that cyclotron motion of ions may be relevant for their specific binding.

• Cyclotron angular velocity does survive room temperatures and the high friction of the thermal bath (water molecules and/or thermal excitations of protein degrees of freedom).

• The orbital moment (3.1) generates diamagnetic moment (3.5), which is small for the considered weak magnetic fields. Hence, our effects are *not* about macroscopic magnetization. However, the diamagnetic moment (3.5) can be observed at large magnetic fields (10 T), if the ion concentration and the white-noise intensity are sufficiently large. This can validate our finding on systems that are experimentally manageable, but artificial from the biological viewpoint.

• Once the non-zero diamagnetic moment relies on the memory of equilibrium thermal bath, we have shown how the memory influences the Brownian motion of the ion. This memory leads to specific underdamping, where (despite a large friction), the ion moves along rotating trajectories for times shorter than the relaxation time. This motion randomizes for times comparable with the relaxation times. A weak magnetic field biases those rotations in a certain direction on much longer times and leads to an average motion cyclotron with the cyclotron frequency.

• It is possible to keep a non-zero magnetic moment also for an external noise with a finite characteristic frequency (infinite frequency corresponds to the white noise); see Appendix C. Then the magnetic moment can be paramagnetic and it nullifies together with the noise frequency.

**5.2 Discussion**
A quasistatic magnetic field is not screened by biological matter (in contrast to electric field) and hence penetrates into an organism without changing its shape and magnitude. This is one reason why magnetic fields apply in biology and medicine. However, mechanisms due to which the static magnetic field influences bio-systems are not clear [6, 7, 8, 9]. One of few cases, where the influence mechanism was clarified includes magnetotaxis phenomena, where certain animal taxa (bacteria, birds, bats, insects, lobsters, salamanders, turtles) are able to sense the Earth magnetic field via synthesizing magnetic particles and joining them in clusters [3, 59, 60]. They can respond to weak magnetic fields for the same reason as the compass responds to the Earth magnetic field, *viz.* due to a strong ferromagnetic interaction between the microscopic magnetic moments that is larger than the thermal energy [60]. Magnetic particles have medical applications, e.g. they are used in magnetic hyperthermia for a local temperature increase [61]. There are more subtle proposals for controlling biological processes (e.g. ion channel functioning) via magnetic particles and external magnetic fields; see [62] and also [63] for a critical review.

Another situation, where a well-known physical effect is applied to magnetobiology includes (bio)chemical reactions with radical pairs [64, 65, 66, 67, 68]. Here the reaction rate depends on spins of reacting radicals, whose coherent (frictionless and noiseless) quantum dynamics involves quasi-degenerate energy levels and hence may be susceptible to weak magnetic fields. The necessary condition for this effect is that the environmental influences (dephasing and relaxation) acting on spins are negligible for reaction times. The radical pair mechanisms are investigated in the context of several scenarios of animal magnetoreception [69, 70]. These mechanisms were so far established *in vitro* only [66], and they also suffer from fragility and non-reproducibility [64, 65]. They cannot apply for describing all scenarios, where weak magnetic fields are relevant for biology [6, 7].

Magnetobiological experiments motivate us to ask how a weak magnetic field can influence ions in a warm and wet cellular environment, i.e., at room temperatures and high friction [6, 7, 8, 9, 22, 24]. So far, there are no theoretical mechanisms that can explain how the influence of a weak, static magnetic field on ions can survive room temperatures and high friction. In particular, the influence of weak magnetic fields is absent in two approaches that are frequently applied for describing cellular physics. First, the high friction nullifies the influence of the magnetic field in the free diffusion. Second, the Bohr-van Leeuwen theorem leads to zero response to a static magnetic field in any (classical) equilibrium system of confined charges [14, 15, 17, 18, 19, 16, 20]. The theorem requires that the coordinate-velocity stationary probability depends on its arguments via the energy only (e.g., Gibbs distribution). Then the theorem rigorously follows from the fact that the magnetic field does not do work, i.e., it does not show up in the energy. On the other hand, the ion motion in a wet and warm cellular environment is certainly classical. Hence all classical stationary state that depends on the magnetic field have to be non-equilibrium.

Our results provide a mechanism that can explain how a weak (static, homogeneous) magnetic field can influence cellular motion of ions. Above assumptions are necessary for obtaining a magnetic response that survives high friction and room temperatures. The first assumption is necessary, because if the *equilibrium* noise is white–and hence the friction is local (Ohmic)–then the external white noise does not violate the Fluctuation-Dissipation Relation (FDR), i.e., the magnetic response is zero thanks to the Bohr-van Leeuwen theorem. Likewise, if we assume that the equilibrium noise is white, but the external noise has a correlation time larger than other characteristic times, then FDR is violated, but still the magnetic response is smaller than (3.1) due to a large friction of the thermal bath; see Appendix C. The assumption on a sufficiently tight confining potential is necessary. Otherwise, the rotational motion of ion will be absent; cf. (3.6). Finally, the assumption on the non-equilibrium noise is also necessary, otherwise the static magnetic field will not influence on the orbital motion of the ion again due to the Bohr-van Leeuwen theorem. Estimating the magnetic moment of ions without friction and/or thermal noise [24] is not relevant for the cellular environment.

The assumption on the white (zero correlation time) character of the non-equilibrium noise is reasonable for an ion interacting with fast protein degrees of freedom, e.g., side-chains rotamers and methyl groups [31], quadrupole moments of $\pi$-covalent bonds within the cation-$\pi$ interaction scenario [71] *etc*. Characteristic times of such degrees of freedom varies between $10^{-15}$ to $10^{-12}$ s, and they are normally described as an external noise acting on ions [57]. Such a fast noise normally does not correlate with slower protein degrees of freedom [47], hence we assumed that random forces in (2.1) do not correlate.

Predictions of our approach can be checked on artificial systems, where various parameters of the model can be tuned. Artificial ion channels [72, 73, 74] are possible candidates for this, since they can provide a sufficiently good confinement of ions inside of pores, together with the possibility to tune (change) features of the solvent. One may envisage that in such systems both the white noise magnitude and the magnetic field can be taken sufficiently large for our effect to be observable via the orbital momentum generated magnetic moment; see Appendix A.

Several pertinent topics are left for future work. *(1)* We shall need a clear theory for fluctuations in stochastic cyclotron motion; see Appendix C for initial steps. *(2)* Since the prediction for the cyclotron motion of cellular ions relates to non-equilibrium, the non-reproducibility of certain magnetobiological experiments can be possibly explained via meta-stability (fragility) of non-equilibrium states. *(3)* Once there is a possibility for a cyclotron motion in a high-friction and high-temperature cellular environment, we shall research on time-dependent electromagnetic fields and possible resonance phenomena. This research will be facilitated if the memory parameter $\kappa$ can be smaller than its value in (2.8). Then the relaxation times obtained from (4.3) will increase sizably, since their dependence on $\kappa$ is non-linear. *(4)* Another pertinent topic is to incorporate electrostatic interactions between the ions that are neglected here. Indeed, the Bjerrum length for bound cations is only a few times smaller than the inter-cation distance estimated from the concentration. *(5)* Future research can establish relations of our results with active matter under magnetic field [38, 39, 41] and electromagnetic noise [17, 43]. *(6)* It will be interesting to generalize this model such that it accounts for a weak magnetic field influence on ion channels [12, 13, 52]. Non-equilibrium noises are abundant for this situation though they are not Gaussian and white [73, 75]. The generalized model has to account for asymmetry, non-linearity and anisotropy of the ion confining potential, selectivity filters *etc*. Ion channels became relevant in the fight against the COVID-19, once the recent literature suggests that its ionic E-channel is critical for its survival [76, 77, 78].

## Acknowledgements

We thank D. Petrosyan and G. Adamian for discussions, and S.G. Gevorkian for explaining us protein physics. Ashot Matevosyan is supported by Cambridge Trust Scholarship. Armen Allahverdyan is supported by SCS of Armenia, grant 20TTAT-QTa003. AA is also partially supported by a research grant from the Yervant Terzian Armenian National Science and Education Fund (ANSEF) based in New York, USA.

## Appendix A. Comparing the non-equilibrium classical and equilibrium quantum average magnetic moments

### Comparison with quantum noise

In the context of (2.1, 2.5), let us emphasize that the very possibility of applying *any* classical Langevin description demands that the characteristic (correlation) of the noise is smaller than the quantum correlation time $k_B T/\hbar$ [79]. For (2.1) this first of all implies

$$\kappa \ll k_B T/\hbar, \tag{A.1}$$

because $\hbar/(k_B T)$ is the characteristic correlation time of the quantum noise [79]. Moreover, the difference in (A.1) should be sufficiently large, so that the correlation frequency of $\boldsymbol{\xi}(t)$ in (2.1), which is was assumed to be much larger than $\kappa$ (i.e., effectively white), also holds (A.1). In (2.8) we estimated $\kappa \sim 10^{10}$ 1/s, while $k_B T/\hbar \sim 10^{13}$ 1/s.

The orbital magnetism announced in (3.1) is a non-equilibrium classical phenomenon. We want to compare it with magnetic effects predicted by equilibrium quantum physics [15, 16]. There are at least two types of such effects: paramagnetism of nuclear spins and diamagnetism of atoms and molecules that is due to the quantum cyclotron motion of electrons bound in atoms.

### Comparison with quantum paramagnetism

Nuclear spins have a tiny equilibrium paramagnetic average magnetic moment even at room temperature [15, 16]. This polarization is experimentally visible at magnetic fields $B \gtrsim 1$T due to a sufficiently large number of identical nuclei. Hence it is interesting to compare the magnitude of this quantum equilibrium effect with the non-equilibrium magnetic moment $\mathcal{M}_c = QL/2$ generated according to (3.1) [cf. (3.5)]:

$$\mathcal{M}_c = -\frac{q_w}{q}\frac{k_B T Q b}{m\kappa^2}, \qquad b = \frac{QB}{m} \tag{A.2}$$

where $Q$ is the elementary charge, and $b$ is the cyclotron frequency.

The interaction energy of the nuclear spin $\frac{1}{2}$ with the magnetic field $B\mathbf{e}_z$ (directed along the z-axis, $\mathbf{e}_z^2 = 1$) is [15, 16]

$$-\hbar\, Q\, B\, \hat{s}_z/(2m), \tag{A.3}$$

where $\hat{s}_z$ is the z-component of the spin-$\frac{1}{2}$ operator (third Pauli's matrix), and $m$ is the nucleus mass, which is taken to be equal to ion's mass in (2.1) (they anyhow have the same order of magnitude), and where $\hbar|Q|/(2m)$ is Bohr's magneton. The $g$-factor in (A.3) is put to 1, since we aim at qualitative estimates. Hence $\widehat{\mathcal{M}} = \hbar Q \hat{s}_z/(2m)$ is the magnetic moment operator [16]. Its equilibrium average value is

$$\langle \widehat{\mathcal{M}} \rangle = \frac{\hbar Q}{2m} \tanh\left[\frac{\hbar Q B}{4m k_B T}\right] = \frac{\hbar^2 Q b}{8 m k_B T}, \tag{A.4}$$

where we employed $\tanh\left[\frac{\hbar Q B}{4m k_B T}\right] \simeq \frac{\hbar Q B}{4m k_B T}$, which holds due to high (room) temperatures. We now get from (A.2, A.4):

$$\frac{|\mathcal{M}_c|}{\langle \widehat{\mathcal{M}} \rangle} = \frac{q_w}{q} \frac{8}{\kappa^2} \left(\frac{k_B T}{\hbar}\right)^2. \tag{A.5}$$

If now (A.1) holds say by two-three order of magnitude, then $|\mathcal{M}_c|$ exceeds $\langle \widehat{\mathcal{M}} \rangle$ even for $q_w \ll q$. Above we compared the (average) magnetic moment (A.2) with the average spin moment (A.4) at the same temperature. This is the most meaningful comparison, but we can also compare (A.2) with the spin magnetic moment (A.4) at $T = 0$, where (A.4) reduces to the (nuclear) Bohr magneton $\langle \widehat{\mathcal{M}} \rangle = \hbar Q/(2m)$:

$$\frac{|\mathcal{M}_c|}{\langle \widehat{\mathcal{M}} \rangle (T=0)} = \frac{q_w}{q} \frac{2 b}{\kappa^2} \left(\frac{k_B T}{\hbar}\right),$$

For parameters from (2.8), we estimate this ratio as $\sim \frac{q_w}{q} \times 10^{-3}$.

**Comparison with quantum diamagnetism**

Turning to quantum diamagnetism, let us quote the standard Langevin's formula for the average magnetic moment of a molecule (atom) that is created by the external magnetic field via induced cyclotron motion of its electrons [14, 15, 16]:

$$\langle \mathcal{M} \rangle_d = -\frac{Z Q^2 r_i^2 B}{6 m_e}, \tag{A.6}$$

where $r_i$ is molecule's radius, $Q$ is the electron charge, $m_e$ is the electron mass, and $Z$ is the number of electrons in the molecule. Note that albeit (A.6) is a quantum effect (we recall that classically there is no equilibrium magnetism) it does not contain $\hbar$ explicitly [14, 15, 16]. It also does not contain any direct reference to the temperature, since it refers to the ground state motion of electrons inside of the molecule.

Let us now see that (A.6) correctly reproduces the diamagnetic volume magnetic susceptibility of water (i.e., the main substance of any cell). To this end, note that (A.6) leads to the macroscopic magnetic moment $\chi_d B/\mu_0$ [15], with susceptibility:

$$\chi_d = -\frac{N_w \mu_0 Z Q^2 r_i^2}{6 m_e}, \tag{A.7}$$

where $\mu_0 = 1.3 \times 10^{-6}$ and $N_w$ is the number of water molecules in meter$^3$. In normal conditions $N_w = \frac{1}{3} \times 10^{29}$ meter$^{-3}$, (this comes from the water density 1g/cm$^3$ and the molar mass of water 18 g/mol), $r_i = 2 \times 10^{-10}$ m, $Q = 1.6 \times 10^{-19}$ Cl, $Z = 10$, and $m_e = 9 \times 10^{-31}$ kg. Putting these numbers into (A.7) we obtain:

$$\chi_d = -7.4 \times 10^{-5}, \tag{A.8}$$

which is close but still larger by its absolute value than the standard value $-9 \times 10^{-6}$ for water [3]. This discrepancy can be explained by noting that there is a (positive!) paramagnetic contribution to be added to (A.8). Overall, the water diamagnetism is explained by (A.6).

We compare (A.2) with (A.6):

$$\frac{\mathcal{M}_c}{\langle \widehat{\mathcal{M}} \rangle_d} = \frac{q_w}{q} \frac{6}{Z} \frac{k_B T m_e}{m^2 r_i^2 \kappa^2}. \tag{A.9}$$

Putting into (A.9) standard estimates for the ion (room temperature, $r_i = 2 \times 10^{-8}$ cm, $m = 3 \times 10^{-23}$ g, $m_e = 9 \times 10^{-28}$ g, $\kappa \sim 10^{10}$ s$^{-1}$, $Z \sim 10$), we find that

$$\frac{\mathcal{M}_c}{\langle \widehat{\mathcal{M}} \rangle_d} \sim \frac{q_w}{q} \frac{6}{Z}, \tag{A.10}$$

i.e., $\mathcal{M}_c$ and $\langle \widehat{\mathcal{M}} \rangle_d$ have the same order of magnitude for $q \sim q_w$. The macroscopic magnetic moment is larger for water, since its concentration is larger. Indeed, the concentration of e.g., bound potassium (K$^+$) ions in a cell is roughly $10^{25}$ meter$^{-3}$ [11], i.e., some 4 orders of magnitude smaller than the water concentration.

Recall that the physical meaning of $\mathcal{M}_c$ is different from that of $\langle \widehat{\mathcal{M}} \rangle_d$. The latter refers to electron motion inside of atoms. The electron motion is frictionless and noiseless: $\langle \widehat{\mathcal{M}} \rangle_d$ refers to the ground state of an atom with its electrons subject to a self-consistent field [15, 16]. In contrast, $\mathcal{M}_c$ refers to ion rotation on scales $\gtrsim 1$ nm and subject to both strong friction and room-temperature thermal noise; see (2.8).

Finally, let us mention that besides frictionless and noiseless $\langle \widehat{\mathcal{M}} \rangle_d$, there is a theoretical possibility of a dissipative equilibrium quantum diamagnetism for ion's motion. It is described by the quantum Langevin equation with a local friction and with the quantum noise with correlation time (A.1) [42]. This effect is however suppressed by friction [42], i.e., it is small under the strong cellular friction (2.8).

## Appendix B. Fokker-Planck equation

We are interested in long-time (stationary) averages generated by (2.1). For reaching the stationary state, it is necessary and sufficient that the potential $u(\mathbf{x})$ is confining, i.e., it goes to infinity for $|\mathbf{x}| \to \infty$ [28, 37]. Moving from (2.1) to the Fokker-Planck equation is not straightforward, since (2.1) is an integro-differential equation and contains the non-white noise $\boldsymbol{\eta}(t)$. To facilitate the construction of the Fokker-Planck equation below, we represent the friction in (2.1) via a linear, first-order in time equation that contains an additional variable **r**. To this end, define

$$\mathbf{r} \equiv \gamma \int_{t_0}^{t} dt'\, \kappa\, e^{-\kappa|t-t'|}\, \mathbf{v}(t'), \tag{B.1}$$

$$\dot{r}_s = -\kappa\, r_s + \gamma\kappa\, v_s, \qquad s = x, y, z. \tag{B.2}$$

Solving (B.2) with initial conditions at $t = t_0$ and taking $t_0 \to -\infty$ we get back (B.1). The limit $t_0 \to -\infty$ is a natural one, since we are eventually interested in the stationary state.

Likewise, the noise $\eta_s$ ($s = x, y, z$) in (2.4) is represented for $t_0 \to -\infty$ as a solution of a linear, first-order in time differential equation with white noise $\varepsilon_s(t)$:

$$\dot{\eta}_s = -\theta\eta_s + \theta\sqrt{q}\, \varepsilon_s(t), \qquad \langle \varepsilon_s(t) \rangle = 0, \qquad \langle \varepsilon_s(t)\varepsilon_{s'}(t') \rangle = \delta_{ss'}\delta(t - t'). \tag{B.3}$$

We define $b = \frac{QB}{m}$ as the cyclotron frequency. For simplicity take $m = 1$ and $Q = 1$. To recover the solution containing mass ($m$) and charge ($Q$), replace in formulas below

$$q \to \frac{q}{m^2}, \qquad q_w \to \frac{q_w}{m^2}, \qquad b \to \frac{QB}{m}. \tag{B.4}$$

Together with the new variables introduced in (B.2, B.3), the Langevin equation (2.1) reads after putting $\mathbf{B} = b\mathbf{e}_z$ (the z-axis is chosen along the magnetic field) and using (B.3, B.1):

$$\dot{\mathbf{v}} = -\mathbf{r} + b\, \mathbf{v} \times \mathbf{e}_z - u_\mathbf{x} + \boldsymbol{\eta} + \boldsymbol{\xi}, \tag{B.5}$$

Projecting (B.5) on the $(x, y, z)$ plane we find

$$\begin{aligned}
\dot{v}_x &= -r_x + b\, v_y - u_x + \eta_x + \xi_x, \\
\dot{v}_y &= -r_y - b\, v_x - u_y + \eta_y + \xi_y, \\
\dot{v}_z &= -r_z - u_z + \eta_z + \xi_z.
\end{aligned} \tag{B.6}$$

Eqs. (B.2, B.3, B.6) are differential, first-order in time and contain only white noises $\varepsilon_s$ and $\xi_s$. Hence for the joint probability $P(\{s, v_s, r_s, \eta_s : s = x, y, z\}; t)$, that now contains additional variables $r_s$ and $\eta_s$, we obtain the Fokker-Planck equation following the standard recipe (see (B.10, B.11) [37]):

$$\begin{aligned}
\partial_t P = \sum_{s=x,y,z} &\left\{ -v_s\, \partial_s P + (r_s + u_s - \eta_s)\, \partial_{v_s} P + \partial_{r_s}[(\kappa r_s - \gamma\kappa v_s)P] \right. \\
&\left. + \theta\, \partial_{\eta_s}(\eta_s P) + \frac{1}{2}q\theta^2\, \partial_{\eta_s}^2 P + \frac{1}{2}q_w\, \partial_{v_s}^2 P \right\} + b(v_x\, \partial_{v_y} - v_y\, \partial_{v_x})P.
\end{aligned} \tag{B.7}$$

For general (confining, but not necessarily harmonic) potentials $u(\mathbf{x})$, eq. (B.7) is not solvable even in the stationary limit $t \to \infty$, where $\partial_t P = 0$. In this limit, (B.7) is solvable only when FDR is valid *globally*, e.g., (2.5) holds and also $q_w = 0$. Then (B.7) produces for $t \to \infty$ the Gibbs distribution for the joint probability of velocity $\mathbf{v}$ and coordinate $\mathbf{x}$ [28, 37, 79]:

$$P(\mathbf{v}, \mathbf{x}) = \frac{1}{Z} e^{-\frac{m}{k_B T}\left(\frac{\mathbf{v}^2}{2} + u(\mathbf{x})\right)}, \quad Z = \int d\mathbf{v} d\mathbf{x}\, e^{-\frac{m}{k_B T}\left(\frac{\mathbf{v}^2}{2} + u(\mathbf{x})\right)}. \tag{B.8}$$

Note that (B.8) does not contain the magnetic field, because the latter does not appear in the energy $m\left(\frac{\mathbf{v}^2}{2} + u(\mathbf{x})\right)$ that determines the Gibbs density. This is the Bohr-van Leeuwen theorem [14, 15, 16]; see [17, 18, 20] for recent discussions. Eq. (B.8) is confirmed below.

### Derivation of the Fokker-Planck equation

First we recall the recipe of writing down the Fokker-Plank equation from the Langevin equation. Consider the following white-noise Langevin equations (assuming summation over repeated indices):

$$\frac{d}{dt}\xi_i = h_i(\xi, t) + g_{ij}\Gamma_j(t), \quad \langle \Gamma_i(t)\, \Gamma_j(t') \rangle = \delta_{ij}\delta(t - t'). \tag{B.9}$$

Then the corresponding Fokker-Plank equation reads [28, 37]

$$\frac{\partial}{\partial t}P = -\frac{\partial}{\partial x_i}\left(D_i^{(1)}(x,t)\, P\right) + \frac{\partial^2}{\partial x_i\, \partial x_j}\left(D_{ij}^{(2)}\, P\right), \tag{B.10}$$

$$D_{ij}^{(2)} = \frac{1}{2}\, g_{ik}\, g_{jk}, \quad D_i^{(1)}(x,t) = h_i(x,t). \tag{B.11}$$

Using (B.10, B.11) we get (B.7) from (B.2, B.3, B.6).

## Appendix C. Equations of moments

### Solving the Fokker-Planck equation via moments

We turn to exact equations for second-order moments. The advantage of this method is that it is able to get a useful information (though not to solve completely) on the case with non-linear potential.

We multiply (B.7) by a function $\psi(x_s, v_s, r_s, \eta_s)$ ($s = x, y, z$) and integrate over all random variables defining the average as $\langle f \rangle = \int d\mathcal{V}\, f\, P$, where $d\mathcal{V} \equiv \prod dx_s dv_s dr_s d\eta_s$. The probability distribution $P$ nullifies at infinity, since the potential $u(\mathbf{x})$ is confining. Hence, after partial integration we get

$$\int d\mathcal{V}\, f\, \partial_\alpha(g\, P) = -\int d\mathcal{V}\, P\, g\, \partial_\alpha f \equiv -\langle g\, \partial_\alpha f \rangle, \tag{C.1}$$

for any functions $f$, $g$ of random variables, where $\alpha$ is one of the variables of integration. Using (C.1) we get for any function $\psi$:

$$\partial_t \langle \psi \rangle = \sum_{s=x,y,z} \{ \langle v_s \, \partial_s \psi \rangle - \langle (r_s + u_s - \eta_s) \, \partial_{v_s} \psi \rangle - \langle (\kappa r_s - \gamma \kappa v_s) \, \partial_{r_s} \psi \rangle$$
$$- \theta \langle \eta_s \, \partial_{\eta_s} \psi \rangle + \tfrac{1}{2} q\theta \langle \partial^2_{\eta_s} \psi \rangle + \tfrac{1}{2} q_w \langle \partial^2_{v_s} \psi \rangle \} - b \langle v_x \, \partial_{v_y} \psi \rangle + b \langle v_y \, \partial_{v_x} \psi \rangle. \quad \text{(C.2)}$$

In the stationary state, the moments are time-independent: $\partial_t \langle \psi \rangle = 0$. We now determine some moments from imposing a natural symmetry on the system.

**Moments constrained by the symmetry**

We now assume that the potential is spherically symmetric

$$u(x, y, z) = u(\sqrt{x^2 + y^2 + z^2}). \quad \text{(C.3)}$$

This assumption holds well for ions bound in (next to) proteins [23, 24]. Our system is rotation-symmetric in the $(x, y)$ plane perpendicular to the magnetic field, since all the forces in (B.5) (besides the magnetic field term) are spherically symmetric. This means that $f(A_x, B_x)$ and $f(A_y, B_y)$ have same statistics for any function $f$ and random variables $A_i, B_i \in \{s, v_s, \eta_s, r_s; s = x, y\}$. Again from the rotational symmetry, rotating system by 90 degrees, the coordinate transformation is $A_x \to A_y$ and $A_y \to -A_x$. Therefore, $\langle f(A_x, B_y) \rangle = \langle f(A_y, -B_x) \rangle$, and

$$\langle A_x B_x - A_y B_y \rangle = 0 \quad \text{and} \quad \langle A_x B_y + A_y B_x \rangle = 0 \quad \text{(C.4)}$$

where $A$, $B$ can be $x$, $v_y$, $\eta_x$, etc. For $f(x, y) = xy$ we have $\langle f(A_x, A_y) \rangle = \langle f(A_y, -A_x) \rangle$ implying that components of the same quantity are uncorrelated: $\langle A_x A_y \rangle = 0$; e.g., $\langle xy \rangle = 0$ and $\langle v_x v_y \rangle = 0$. These results also follow from (C.2).

**Working out the moment equations**

Turning to moments generated via (C.2), let us recall that many moments are equal or cancel each other. Hence we denote:

$$\langle A_x, B_y \rangle_\pm \equiv \langle A_x B_y \pm A_y B_x \rangle,$$
$$\langle A_x, B_x \rangle_\pm \equiv \langle A_x B_x \pm A_y B_y \rangle, \quad \text{(C.5)}$$

where e.g., $\langle x, v_y \rangle_- \equiv \langle x v_y - y v_x \rangle$ or $\langle x, x \rangle_+ = \langle x^2 + y^2 \rangle$.

Using a function $\psi = g(x, y, z)$ in the stationary regime of (C.2) we find

$$0 = \langle v_x \, \partial_x g \rangle + \langle v_y \, \partial_y g \rangle + \langle v_z \, \partial_z g \rangle, \quad \text{(C.6)}$$

hence

$$g = xy: \qquad \langle xv_y\rangle + \langle yv_x\rangle = 0$$
$$g = x^2; y^2; z^2: \qquad \langle xv_x\rangle = \langle yv_y\rangle = \langle zv_z\rangle = 0 \qquad (C.7)$$
$$g = u(x,y,z): \qquad \langle v_x u_x + v_y u_y + v_z u_z\rangle = 0$$

One can easily show that first-order moments, i.e., average values of the random variables are all zero. Then consider $\psi = \eta_x \eta_y, \eta_x^2, \eta_y^2$ in (C.2):

$$\langle \eta_x^2\rangle = \langle \eta_y^2\rangle = \frac{1}{2}q\theta, \qquad \langle \eta_x \eta_y\rangle = 0. \qquad (C.8)$$

Another set of simple relations is found by putting $\psi = \eta_i j$ and $\psi = \eta_i r_j$ $(i,j = x,y)$ into (C.2):

$$\langle v_i \eta_j\rangle = \theta \langle x_i \eta_j\rangle, \qquad (C.9)$$

$$\langle r_i \eta_j\rangle = \gamma\mu \langle x_i \eta_j\rangle, \qquad \mu = \frac{\kappa\theta}{\kappa + \theta}. \qquad (C.10)$$

Similar relations we get from $\psi = r_x^2, r_y^2, xr_x, yr_y$ and $xr_y + yr_x$:

$$\langle v_i r_i\rangle = \kappa\langle x_i r_i\rangle, \qquad \langle v_x, r_y\rangle_+ = \kappa\langle x, r_y\rangle_+, \qquad \langle r_i^2\rangle = \gamma\kappa\langle x_i r_i\rangle. \qquad (C.11)$$

In the following formulas the left column denotes the function $\psi$ to be employed in (C.2), while the result are reported on the right using notations (C.5):

$$v_x \eta_x + v_y \eta_y: \qquad (\theta^2 + \gamma\mu)\langle x, \eta_x\rangle_+ + \langle u_x, \eta_x\rangle_+ + b\theta\langle x, \eta_y\rangle_- = \theta q \qquad (C.12)$$

$$v_y \eta_x - v_x \eta_y: \qquad (\theta^2 + \gamma\mu)\langle x, \eta_y\rangle_- + \langle u_x, \eta_y\rangle_- - b\theta\langle x, \eta_x\rangle_+ = 0 \qquad (C.13)$$

$$v_x^2 + v_y^2: \qquad \kappa\langle x, r_x\rangle_+ = \theta\langle x, \eta_x\rangle_+ + (\langle u_z v_z\rangle + q_w) \qquad (C.14)$$

$$xv_y - yv_x: \qquad \langle x, r_y\rangle_- = \langle x, \eta_y\rangle_- \qquad (C.15)$$

$$v_x r_y - v_y r_x: \qquad \gamma\mu\langle x, r_y\rangle_- + \langle u_x, r_y\rangle_- + \kappa\langle v_x, r_y\rangle_- = b\kappa\langle x, r_x\rangle_+, \qquad (C.16)$$

where in (C.16) we used (C.10). This system of equations is closed (assuming expressions with $u$ are known). Note that $\langle x, \eta_x\rangle_+$ and $\langle x, \eta_y\rangle_-$ are interesting quantities that appear in intermediate steps of the solution: they refer to (resp.) power and torque of the noise. Likewise, $\langle x, r_x\rangle_+$ and $\langle x, r_y\rangle_-$ refers to the friction force. The next three equations contain the quantities that we are interested in:

$$xr_y - yr_x: \qquad \gamma\kappa\langle x, v_y\rangle_- = -\langle v_x, r_y\rangle_- + \kappa\langle x, r_y\rangle_- \qquad (C.17)$$

$$r_x v_x + r_y v_y: \qquad \gamma\kappa\langle v_x, v_x\rangle_+ = (\gamma + \kappa)\kappa\langle x, r_x\rangle_+ + \langle u_x, r_x\rangle_+ \qquad (C.18)$$

$$xv_x + yv_y: \quad \langle x, u_x\rangle_+ = \langle v_x, v_x\rangle_+ + b\langle x, v_y\rangle_- \quad \text{(C.19)}$$

$$\begin{aligned}&-\gamma\mu\langle x,\eta_x\rangle_+ + b\langle v_x,r_y\rangle_-\\&+\langle x,\eta_x\rangle_+ - \langle x,r_x\rangle_+\end{aligned}$$

Now we assume $\theta = \kappa$, i.e., FDR (2.5) holds. Eqs. (C.12–C.19) are solved as

$$\begin{aligned}
\langle x, v_y\rangle_- &= -b\frac{q'_w}{\gamma\kappa^2} + \frac{1}{\gamma\kappa^2}\langle u_x, \eta_y - r_y\rangle_-\\
\langle v_x, v_x\rangle_+ &= \frac{q}{\gamma} + \frac{q'_w}{\gamma}\frac{b^2 + \gamma\kappa + \kappa^2}{\kappa^2} + \frac{b\langle u_x, \eta_y - r_y\rangle_- + \kappa\langle u_x, r_x - \eta_x\rangle_+}{\gamma\kappa^2}\\
\langle x, u_x\rangle_+ &= \frac{q + q'_w}{\gamma} + \frac{\langle u_x, r_x - \eta_x\rangle_+}{\gamma\kappa}, \qquad q'_w \equiv q_w + \langle u_z v_z\rangle.
\end{aligned} \quad \text{(C.20)}$$

This is the farthermost point where we can reach with a general spherical symmetric potential. Assuming that FDR (2.5) holds, we write (C.20) as

$$\langle L\rangle \equiv \langle xv_y - yv_x\rangle = -b\frac{q'_w}{\gamma\kappa^2} + 2\times\frac{1}{\gamma\kappa^2}\langle u_x(\eta_y - r_y)\rangle, \quad \text{(C.21)}$$

$$q'_w \equiv q_w + \langle u_z v_z\rangle, \quad \text{(C.22)}$$

$$\langle v_x^2 + v_y^2\rangle = \frac{q}{\gamma} + \frac{q'_w}{\gamma}\frac{b^2 + \gamma\kappa + \kappa^2}{\kappa^2} + 2\times\frac{b\langle u_x(\eta_y - r_y)\rangle + \kappa\langle u_x(r_x - \eta_x)\rangle}{\gamma\kappa^2}, \quad \text{(C.23)}$$

$$\langle xu_x + yu_y\rangle = \frac{q + q'_w}{\gamma} + 2\times\frac{\langle u_x(r_x - \eta_x)\rangle}{\gamma\kappa}, \quad \text{(C.24)}$$

where $L = xv_y - yv_x$ is the z-component of the orbital momentum (i.e., the component along the magnetic field), and where $\mathbf{r} - \boldsymbol{\eta}$ is the excess force acting on the ion; cf. (B.5).

It is seen from (C.21) that generally $\langle L\rangle \neq 0$. Note that for a harmonic potential $u_x \propto x$ and we get $\langle u_x, \eta_y - r_y\rangle_- = \langle u_x(\eta_y - r_y)\rangle = 0$ in (C.20) and (C.21), as seen from (C.15). Another simplification of this potential is that $\langle u_z v_z\rangle \propto \langle z\, v_z\rangle \propto \left\langle\frac{d}{dt} v_z^2\right\rangle = 0$, and hence $q'_w = q_w$ in (C.21). Hence, we obtain from (C.21) the harmonic potential expression (3.1). Note that for anharmonic potentials the term $\langle u_x, \eta_y - r_y\rangle_-$ does not generally nullify, but the sign of $\langle L\rangle/B$ stays negative, i.e. the response is still diamagnetic, as we checked numerically.

Now $\langle L\rangle$ nullifies for three cases; see (C.21). First, $\langle L\rangle = 0$ when $\kappa \to \infty$ in (C.21), since now we get that for long times the system holds FDR with white noises and memoryless friction. Hence it relaxes to the Gibbs distribution (B.8) that does not contain the magnetic field $b$. This leads to $\langle L\rangle = 0$. In the same limit $\kappa \to \infty$ we get from (C.23) the known thermal expression for $\langle v_x^2 + v_y^2\rangle = 2\langle v_x^2\rangle$. Second, $\langle L\rangle = 0$ for $q_w = 0$ (zero intensity of the white noise) where FDR again holds. Now we should note that $q'_w = \langle u_z v_z \rangle = \langle u_z \rangle \langle v_z \rangle = 0$ in (C.22), because within the (stationary) Gibbs distribution (B.8), the coordinate and momentum factorize and $\langle \mathbf{v} \rangle = 0$. In (C.24) $q/\gamma$ is already the result of the Gibbs distribution, as seen from (B.8, 2.6) upon using (B.4). Hence, (C.24) plus the Gibbs distribution imply $\langle u_x(r_x - \eta_x) \rangle = 0$. For the same reason $\langle u_x(\eta_y - r_y) \rangle = 0$ in (C.23), which leads to $\langle L \rangle = 0$ in (C.21). Third, $\langle L \rangle = 0$ for $b \to 0$ (zero magnetic field), this time because the spherical symmetry holds for the stationary state, which e.g., leads to $\langle u_x \eta_y \rangle = -\langle u_x \eta_y \rangle = 0$ after inverting $x \to -x$.

**Harmonic potential**

In the case of harmonic potential $u = \frac{\omega_0^2}{2}(x^2 + y^2 + z^2)$ the moments that contain potential term become $\langle u_i f_j \rangle = \omega_0^2 \langle x_i f_j \rangle$ for $i, j = x, y, z$. After this replacement, (C.12–C.19) becomes a solvable system of linear equations. The additional factor $\langle u_z v_z \rangle$ in (C.22) nullifies, since for the harmonic potential the motion in the $z$-direction separates from the motion in the $(x, y)$-plane and then $\langle u_z v_z \rangle = \frac{1}{2} \omega_0^2 \left\langle \frac{d(z^2)}{dt} \right\rangle = 0$ in the stationary state.

For more generality, we shall replace the external white noise in (2.4) with correlation function $\langle \xi_i(t) \xi_j(t') \rangle = q_w \delta_{ij} \delta(t - t')$ by a colored noise with correlation function:

$$\langle \xi_i(t) \xi_j(t') \rangle = \delta_{ij}\, q_{\text{ext}}\, \theta_{\text{ext}}/2\; e^{-\theta_{\text{ext}}|t-t'|}. \tag{C.25}$$

Then (without assuming $\kappa = \theta$) we can find the following general formulas:

$$\begin{aligned}
\gamma \langle v_x^2 + v_y^2 \rangle &= q\, \frac{\theta^2}{\kappa^2}\, \frac{(\omega_0^2 + \theta^2 + \gamma\mu)(\omega_0^2 + \kappa^2 + \gamma\mu) + b^2\theta^2}{(\omega_0^2 + \theta^2 + \gamma\mu)^2 + b^2\theta^2} \\
&\quad + q_{ext}\, \frac{\theta_{ext}^2}{\kappa^2}\, \frac{(\omega_0^2 + \theta_{ext}^2 + \gamma\mu_{\text{ext}})(\omega_0^2 + \kappa^2 + \gamma\mu_{\text{ext}}) + b^2\theta_{ext}^2}{(\omega_0^2 + \theta_{ext}^2 + \gamma\mu_{\text{ext}})^2 + b^2\theta_{ext}^2}, \\
\gamma \omega_0^2 \langle x^2 + y^2 \rangle &= q\, \frac{\theta^2}{\kappa^2}\, \frac{(\omega_0^2 + \theta^2 + \gamma\mu)(\omega_0^2 + \kappa^2 + \gamma\mu\, \kappa^2/\theta^2) + b^2\kappa^2}{(\omega_0^2 + \theta^2 + \gamma\mu)^2 + b^2\theta^2} \\
&\quad + q_{ext}\, \frac{\theta_{ext}^2}{\kappa^2}\, \frac{(\omega_0^2 + \theta_{ext}^2 + \gamma\mu_{\text{ext}})(\omega_0^2 + \kappa^2 + \gamma\mu_{\text{ext}}\, \kappa^2/\theta_{ext}^2) + b^2}{(\omega_0^2 + \theta_{ext}^2 + \gamma\mu_{\text{ext}})^2 + b^2\theta_{ext}^2} \\
\gamma \langle L \rangle &= \gamma \langle x v_y - y v_x \rangle = q\, \frac{\theta^2}{\kappa^2}\, \frac{b(\kappa - \theta)(\kappa + \theta)}{(\omega_0^2 + \theta^2 + \gamma\mu)^2 + b^2\theta^2} \\
&\quad + q_{ext}\, \frac{\theta_{ext}^2}{\kappa^2}\, \frac{b(\kappa - \theta_{\text{ext}})(\kappa + \theta_{\text{ext}})}{(\omega_0^2 + \theta_{ext}^2 + \gamma\mu_{\text{ext}})^2 + b^2\theta_{ext}^2}.
\end{aligned} \tag{C.26}$$

where $\mu$ is defined in (C.10), and where

$$\mu_{\text{ext}} = \frac{\kappa \theta_{\text{ext}}}{\kappa + \theta_{\text{ext}}}. \tag{C.27}$$

Now (3.1–3.3) of the main text are found from (C.26) under $\kappa = \theta$, $\theta_{\text{ext}} \to \infty$ and $q_{\text{ext}} = q_w$, i.e., in the white-noise limit for the external noise. Assuming $\kappa = \theta$ in (C.26), we find

$$\langle L \rangle = \frac{q_{\text{ext}} \theta_{\text{ext}}^2}{\gamma \kappa^2} \frac{b(\kappa - \theta_{\text{ext}})(\kappa + \theta_{\text{ext}})}{(\omega_0^2 + \theta_{\text{ext}}^2 + \gamma \mu_{\text{ext}})^2 + b^2 \theta_{\text{ext}}^2}. \tag{C.28}$$

For $\kappa = \theta_{\text{ext}}$ the FDR effectively holds, we are back to the equilibrium situation, hence $\langle L \rangle = 0$, as seen from (C.28). Note that for $\kappa > \theta_{\text{ext}}$, we get from (C.28) a paramagnetic response $\langle L \rangle > 0$, which is a simple consequence of $L|_{\kappa=\theta_{\text{ext}}} = 0$ [19].

**Particular cases**

To understand the influence of $\kappa > \theta$ in (C.26), let us take $\kappa = 2\theta$. We find $\gamma \langle L \rangle = -\frac{q_{\text{ext}}}{\kappa^2} + \frac{27}{16} \frac{q}{\gamma^2}$, i.e. a paramagnetic correction to the diamagnetic contribution.

Next, let us study the limit $\kappa \gg \theta_{\text{ext}}$ and $\gamma \gg \theta_{\text{ext}}$, and also neglect $b^2 \theta_{\text{ext}}^2$ in (C.28):

$$\langle L \rangle = \frac{q_{\text{ext}} \theta_{\text{ext}}^2 b}{\gamma (\omega_0^2 + \gamma \theta_{\text{ext}})^2}. \tag{C.29}$$

Note from (C.29) that $H$ tends to zero together with $\theta_{\text{ext}}$. For an ion in shallow potential with a sufficiently small $\omega_0$, there can be an intermediate regime, where $\omega_0^2 \ll \gamma \theta_{\text{ext}}$. Hence we get

$$\langle L \rangle = b q_{\text{ext}} / \gamma^3. \tag{C.30}$$

Comparing this formula with (3.1), we see that under natural condition $q_{\text{ext}} \sim q_w$, (C.) is much smaller than the absolute value of (3.1); cf. (2.8). Note that (C.30) holds qualitatively also when $\theta_{\text{ext}} \lesssim \kappa$; e.g., for $\theta_{\text{ext}} = \kappa/2$ we get from (C.28):

$$\langle L \rangle = \frac{3 q_{\text{ext}} \kappa^2 b}{16 \gamma (\omega_0^2 + \frac{\kappa^4}{4} + \frac{\kappa \gamma}{3})^2} \leq \frac{27 q_{\text{ext}} b}{16 \gamma^3}. \tag{C.31}$$

Let us finally note from (C.) the form of $\langle x^2 + y^2 \rangle$ for $\kappa \gg \theta_{\text{ext}}$ and negligible $b^2 \kappa^2$ and $b^2 \theta_{\text{ext}}^2$:

$$\gamma \omega_0^2 \langle x^2 + y^2 \rangle = q + q_{\text{ext}} \frac{\frac{\omega_0^2}{\kappa^2} + 1 + \frac{\gamma}{\theta_{\text{ext}}}}{\frac{\omega_0^2}{\theta_{\text{ext}}^2} + 1 + \frac{\gamma}{\theta_{\text{ext}}}} \leq q + q_{\text{ext}}. \tag{C.32}$$

# Appendix D. Mean and variance of angular velocity

For the harmonic potential (2.2) the joint probability density for the coordinates and velocities is Gaussian, as seen e.g., from the fact that (2.1) is a linear equation with Gaussian noises [37]. Introducing a column vector $\boldsymbol{\phi} = (v_x, v_y, x, y)^{\text{T}}$ (transposed row vector), we get from (3.1–3.3) for this joint density:

$$P(\boldsymbol{\phi}) = \frac{1}{(2\pi)^2\sqrt{\det \Sigma}} \exp\left[-\frac{1}{2}\boldsymbol{\phi}^T \Sigma^{-1} \boldsymbol{\phi}\right],$$

$$\Sigma = \begin{pmatrix} S & 0 & 0 & -H \\ 0 & S & H & 0 \\ 0 & H & R & 0 \\ -H & 0 & 0 & R \end{pmatrix}, \quad \Lambda = \Sigma^{-1} = \beta \begin{pmatrix} R & 0 & 0 & H \\ 0 & R & -H & 0 \\ 0 & -H & S & 0 \\ H & 0 & 0 & S \end{pmatrix},$$

where $2H = \langle xv_y - yv_x \rangle$, $2S \equiv \langle v_x^2 + v_y^2 \rangle$, $2R \equiv \langle x^2 + y^2 \rangle$, and $\beta \equiv (SR - H^2)^{-1}$. Then probability density (D.1) reads

$$P(\phi) \propto \exp\left[-\frac{1}{2}\left(\beta R(v_x^2 + v_y^2) + \beta S(x^2 + y^2)\right) + \beta H(xv_y - yv_x)\right].$$

From this equation we calculate the marginal density $P(x, y)$ and conditional density $P(v_x, v_y \mid x, y)$:

$$P(x, y) \propto \exp\left[-\frac{x^2 + y^2}{2R}\right] \tag{D.1}$$

$$P(v_x, v_y \mid x, y) = \frac{P(\phi)}{P(x, y)} \propto \exp\left[-\frac{\beta R}{2}(v_x^2 + v_y^2) + \beta H(xv_y - yv_x)\right] \tag{D.2}$$

$$\propto \exp\left[-\frac{\beta R}{2}\left(v_x + \frac{Hy}{R}\right)^2 - \frac{\beta R}{2}\left(v_y - \frac{Hx}{R}\right)^2\right]. \tag{D.3}$$

Equations (D.1–D.3) imply [$E(X) \equiv \langle X \rangle$, and $E(X|Y)$ means the conditional average]:

$$3E[v_x \mid x, y] = -\frac{Hy}{R} \quad \text{and} \quad E[v_y \mid x, y] = \frac{Hx}{R} \tag{D.4}$$

$$E[v_x^2 \mid x, y] = \left(\frac{Hy}{R}\right)^2 + \frac{1}{\beta R} \quad \text{and} \quad E[v_y^2 \mid x, y] = \left(\frac{Hx}{R}\right)^2 + \frac{1}{\beta R}. \tag{D.5}$$

**Mean and variance of the angular velocity $\Omega$**

We define the angular velocity in the $(x, y)$-plane:

$$\Omega = \frac{xv_y - yv_x}{x^2 + y^2}, \tag{D.6}$$

and look at its average:

$$\mathrm{E}[\Omega] = \mathrm{E}[\mathrm{E}[\Omega \mid x, y]] = \mathrm{E}\left[\mathrm{E}\left[\frac{xv_y - yv_x}{x^2 + y^2} \mid x, y\right]\right] \quad (D.7)$$

$$= \mathrm{E}\left[\frac{x}{x^2 + y^2}\mathrm{E}[v_y \mid x, y] - \frac{y}{x^2 + y^2}\mathrm{E}[v_x \mid x, y]\right] \quad (D.8)$$

$$= \mathrm{E}\left[\frac{x}{x^2 + y^2}\frac{Hx}{R} + \frac{y}{x^2 + y^2}\frac{Hy}{R}\right] = \frac{H}{R} \quad (D.9)$$

Note that $\Omega$ has a singularity at $\rho \equiv x^2 + y^2 = 0$. Thus we study the statistics of $\Omega$ at a fixed $\rho$. From (D.9) it is easy to see that $\mathrm{E}[\Omega \mid \rho] = \mathrm{E}[\Omega]$. Next, we calculate

$$\mathrm{E}[\Omega^2 \mid \rho] = \mathrm{E}\left[\frac{x^2 v_y^2 + y^2 v_x^2 - 2xy v_x v_y}{(x^2 + y^2)^2} \mid \rho\right] \quad (D.10)$$

$$= \mathrm{E}\left[\frac{x^2\,\mathrm{E}[v_y^2 \mid x, y]}{(x^2 + y^2)^2} + \frac{y^2\,\mathrm{E}[v_x^2 \mid x, y]}{(x^2 + y^2)^2} - \frac{2xy\,\mathrm{E}[v_x v_y \mid x, y]}{(x^2 + y^2)^2} \mid \rho\right] \quad (D.11)$$

$$= \mathrm{E}\left[\frac{x^2 + y^2}{(x^2 + y^2)^2}\frac{1}{\beta R} + \frac{x^4 + y^4}{(x^2 + y^2)^2}\left(\frac{H}{R}\right)^2 + \frac{2x^2 y^2}{(x^2 + y^2)^2}\left(\frac{H}{R}\right)^2 \mid \rho\right] \quad (D.12)$$

$$= \frac{RS - H^2}{R}\frac{1}{\rho^2} + \frac{H^2}{R^2}, \quad (D.13)$$

and thus

$$\langle\Omega\rangle = \frac{H}{R}, \qquad \mathrm{Var}[\Omega \mid \rho] = \frac{1}{\rho^2}\frac{RS - H^2}{R} \quad (D.14)$$

It is seen that $\langle\Omega^2\rangle$ does not exist due to a divergence induced by the factor $\frac{1}{\rho^2}$. Indeed, the calculation of $\langle\Omega^2\rangle$ involves the integral $\propto \int_0 \mathrm{d}\rho\, \rho\, \rho^{-2}$, which diverges logarithmically.

# Appendix E. Calculation of auto-correlation functions

Since the Fokker-Plank equation (B.7) does not account for autocorrelation functions, we will use the Fourier method [16] for solving Langevin's equation (B.5) directly. To make the solution more transparent, we restrict ourselves with the harmonic potential (2.2) and consider only the white noise, i.e., $q = 0$ and $q_w > 0$ in (B.5). These restrictions suffice for having a well-defined average angular motion induced by the magnetic field; cf. (3.1). Note as well that $q = 0$ does not alter the time-dependence of autocorrelation fuction for the harmonic potential.

Hence we focus on $x$ and $y$ coordinates of (2.1) that read together

$$\zeta \equiv x + iy, \qquad \xi \equiv \xi_x + i\xi_y, \quad (E.1)$$

$$\ddot{\zeta}(t) = \xi(t) - \omega_0^2\,\zeta - ib\dot{\zeta}(t) - \gamma \int_{-\infty}^{t} \mathrm{d}t'\, \kappa\, e^{-\kappa|t-t'|}\, \dot{\zeta}(t'), \quad (E.2)$$

where in (2.1) we took $t_0 = -\infty$ aiming to focus on the stationary state. We took $m = 1$ in (E.2).

Equation (E.2) belongs to the class of linear oscillators driven subject to non-local friction and noise. Such systems were studied in [18, 19, 21, 38, 39, 40, 44, 41, 80]. We apply Fourier's transform to $\zeta$ and $\xi$ in (E.2):

$$X(t) = \int d\omega \, e^{i\omega t} \tilde{X}(\omega), \qquad \tilde{X}(\omega) = \int \frac{dt}{2\pi} e^{-i\omega t} X(t), \tag{E.3}$$

where $X(t) = (\zeta(t), \xi(t))$ and $\tilde{X}(\omega) = (\tilde{\zeta}(\omega), \tilde{\xi}(\omega))$, and get from (E.2):

$$\tilde{\zeta}(\omega) = \frac{\tilde{\xi}(\omega)}{i\omega(i\omega + \frac{\gamma \kappa}{\kappa + i\omega} + ib) + \omega_0^2}. \tag{E.4}$$

Now (2.4, E.1) lead to $\langle \xi(t_1)\xi^*(t_2) \rangle = 2q_w \delta(t_1 - t_2)$. This implies $\langle \tilde{\xi}(\omega_1)\tilde{\xi}^*(\omega_2) \rangle = \frac{q_w}{\pi}\delta(\omega_1 - \omega_2)$ from (E.3). Hence we find from (E.4):

$$\langle \zeta(t_1)\zeta^*(t_2) \rangle = \int d\omega_1 \, d\omega_2 \, e^{i\omega_1 t_1 - i\omega_2 t_2} \langle \tilde{\zeta}(\omega_1)\tilde{\zeta}^*(\omega_2) \rangle \tag{E.5}$$

$$= \frac{q_w}{\pi} \int d\omega \, e^{i\omega(t_1 - t_2)} \frac{|\omega - i\kappa|^2}{|(\omega - i\kappa)(\omega^2 + \omega b - \omega_0^2) - \gamma\kappa\omega|^2}. \tag{E.6}$$

For $t_1 - t_2 > 0$ only the poles with positive imaginary parts contribute into (E.6), since the integration contour can be closed above the real axis (in total, (E.6) has 6 simple poles). It turns out that those poles $\omega_1$, $\omega_2$ and $\omega_3$ (with imaginary parts) are the roots of a cubic equation

$$(\omega - i\kappa)(\omega^2 + \omega b - \omega_0^2) - \gamma\kappa\omega = 0. \tag{E.7}$$

Equation (E.7) produces (4.3) after introducing $\omega \equiv iu$. To show that all roots of (4.3) have positive real parts we can employ the generalized Routh-Hurwitz criterion [81]. Recall that the usual Routh-Hurwitz criterion applies to polynomials with real coefficients [82].

# Figures

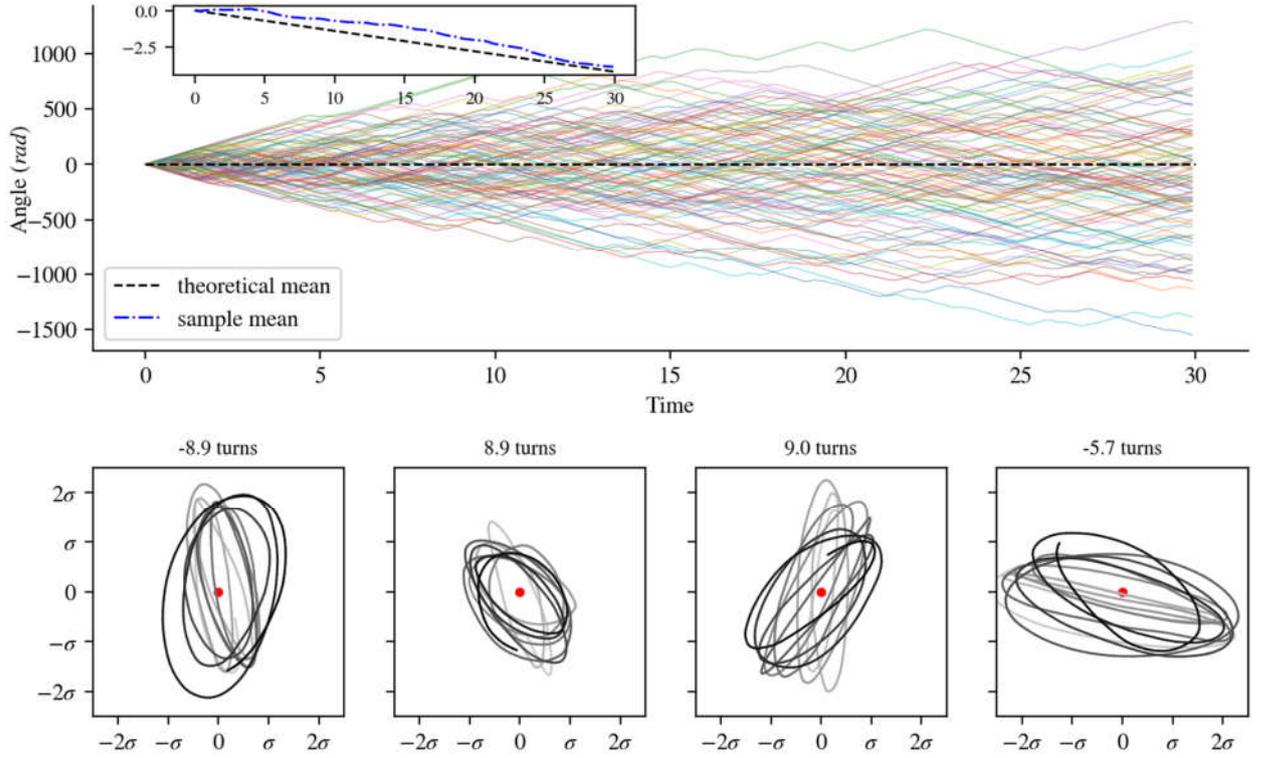

**Figure 1.** (No color in print) The figure shows stochastic trajectories obtained from solving the $x$ and $y$ components of Langevin's equation (2.1) (see also (B.2, B.3, B.6)) for the harmonic potential (2.2). It is seen that despite a strong friction, the random motion of the ion in a confining potential involves rotating trajectories, if the friction has memory (underdamping due to memory). The parameters are: $b = 1$, $\gamma = 100$, $\kappa = 10$, $\omega_0 = 90$, $q_w = 0.1$, $q = 1$, $\sigma \equiv \sqrt{\langle x^2 \rangle} = \sqrt{\langle y^2 \rangle}$; see (2.3) and (3.3).

Top row: the angle $\int_0^t ds\, \Omega(s)$ around the origin for 20 different trajectory realizations. The angular velocity $\Omega(t)$ is given by (3.7). Its average is $\langle \Omega \rangle = -0.775$ according to (3.6). Each color refers to a separate realization of the random motion. There are trajectories, where the angular velocity $\Omega$ stays constant for a long time (rotations).

Inset: the thick blue (dot-dashed) line in the inset is the average over 19200 independent realizations and the thick black (dashed) line shows the theoretical result for $\langle \Omega \rangle t$; see (3.6).

Second row: 4 random trajectories in the $(x, y)$ plane; the red point denotes the origin $(0,0)$. For stochastic trajectories, the darker (brighter) points refer to recent (earlier) times. It is seen that the trajectories resemble deterministic rotations due to the memory and despite a strong friction.

The relaxation times and rotational periods are (resp.) $({u'_1}^{-1}, {u'_2}^{-1}, {u'_3}^{-1}) = (0.1122, 1.8454, 1.8265)$ and $({u''_1}^{-1}, {u''_2}^{-1}, {u''_3}^{-1}) = (-943.6, 0.0104, -0.0105)$; cf. (4.3, 4.2). Thus for $t > 3$ the system is in the stationary state, but rotations are well-visible, since ${u''_2}^{-1} \ll {u'_2}^{-1}$ and $|u''_3|^{-1} \ll {u'_3}^{-1}$ in (4.1). The difference between $u''_2$ and $|u''_3|$ is due to $b \neq 0$.

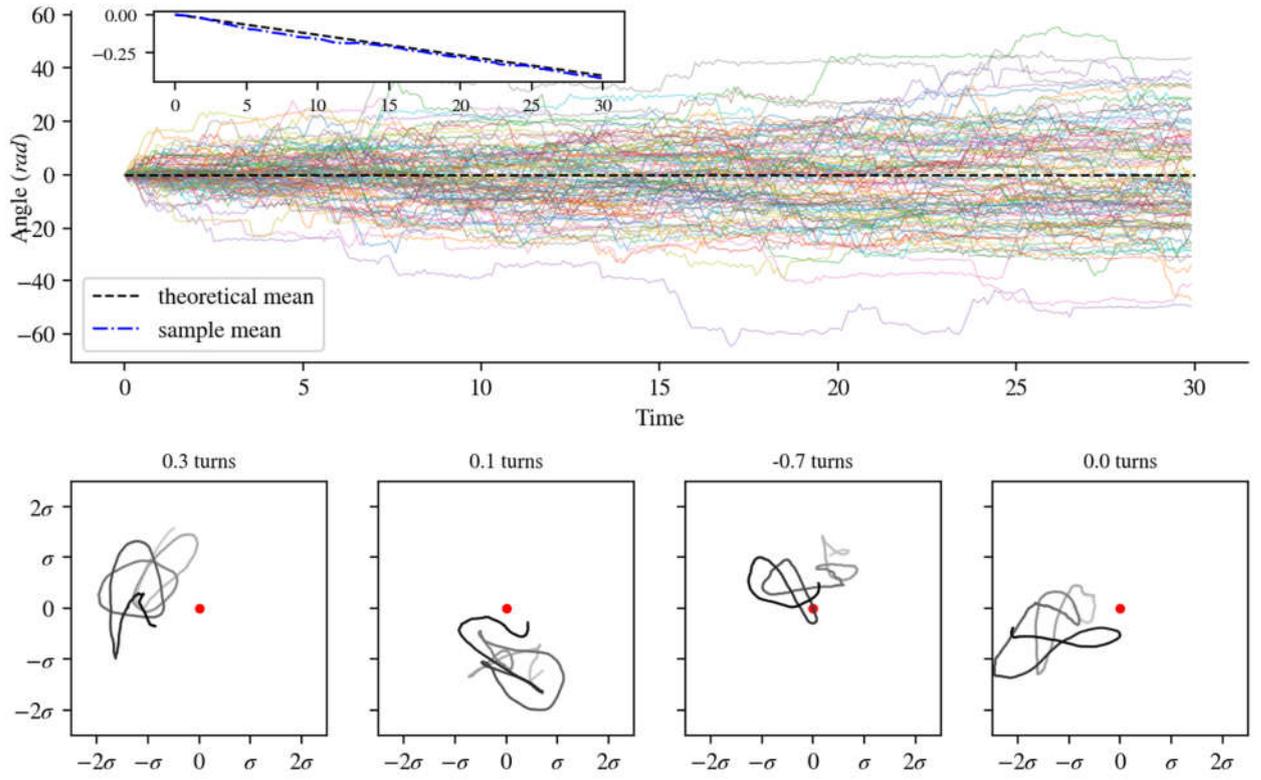

**Figure 2.** (No color in print) The same as in Fig. 1, but for $\omega_0 = 10$, i.e., the potential (2.2) is less confining. It is seen that deterministically rotating trajectories of Fig. 1 are absent, though some signs of rotations are still visible. Now $\langle \Omega \rangle = -0.083$ from (3.6). The relaxation times and rotation periods are (resp.) $({u'}_1^{-1}, {u'}_2^{-1}, {u'}_3^{-1}) = (1.0917, 0.2236, 0.2168)$ and $({u''}_1^{-1}, {u''}_2^{-1}, {u''}_3^{-1}) = (-129.9, 0.030, -0.031)$. The relaxation time ${u'}_2^{-1} \simeq {u'}_3^{-1}$ is still some *10* times larger than the rotation period ${u''}_2^{-1} \simeq |{u'}_3^{-1}|$.

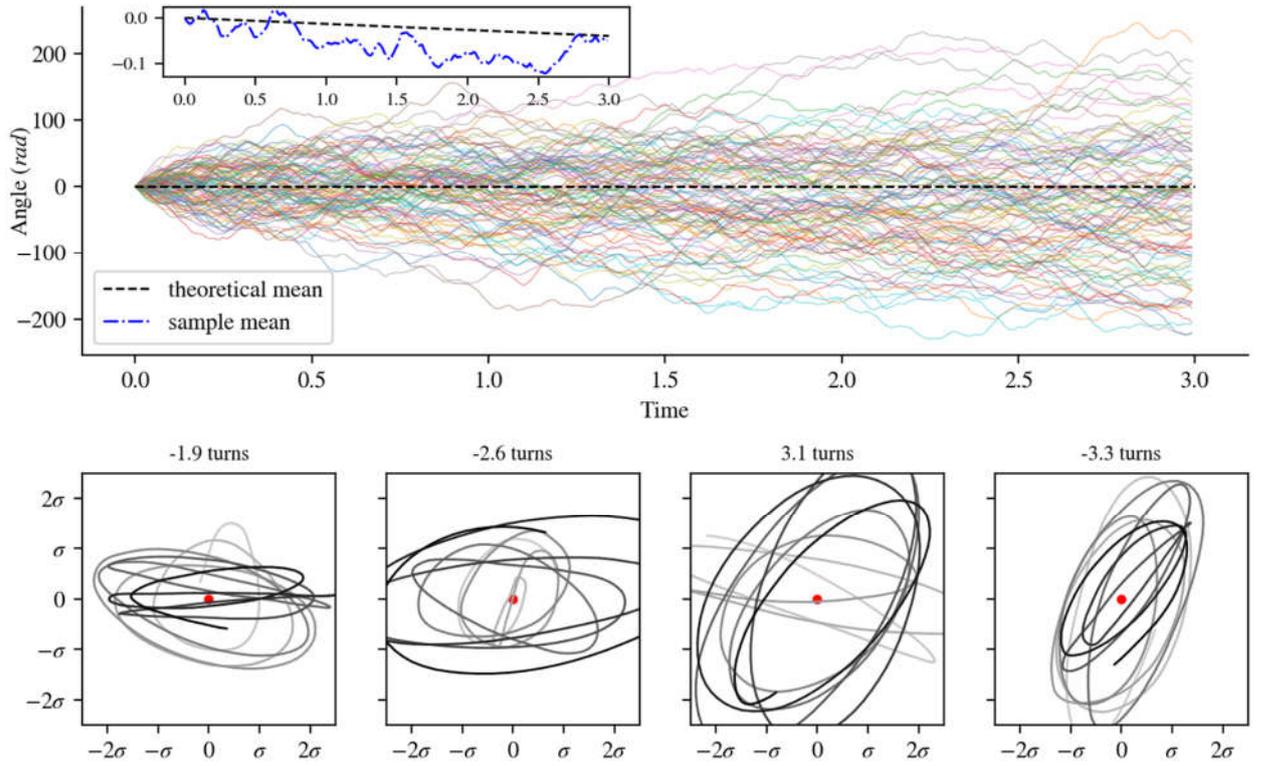

**Figure 3.** (No color in print) The same as in Fig. 1, but with $\kappa = 1000$ and $\omega_0 = 1000$, i.e., the potential is more confining than in Fig. 1, but the memory time is 100 times shorter than in Fig. 1. We see that certain patterns of rotation are clearly visible (cf. Fig. 2), but still they are less deterministic than those in Fig. 1 (underdamping due to a strongly confining potential). Now $\langle \Omega \rangle = -0.083$ from *(3.6)*. The relaxation times and rotation periods are (resp.): $(u'^{-1}_1, u'^{-1}_2, u'^{-1}_3) = (0.001, 0.040, 0.040)$ and $(u''^{-1}_1, u''^{-1}_2, u''^{-1}_3) = (-40.16, 0.00097, -0.00097)$. The shortest rotation periods are much smaller than the relaxation times: $u''^{-1}_2 \simeq |u''_3|^{-1} \ll u'^{-1}_2 \simeq u'^{-1}_3$.

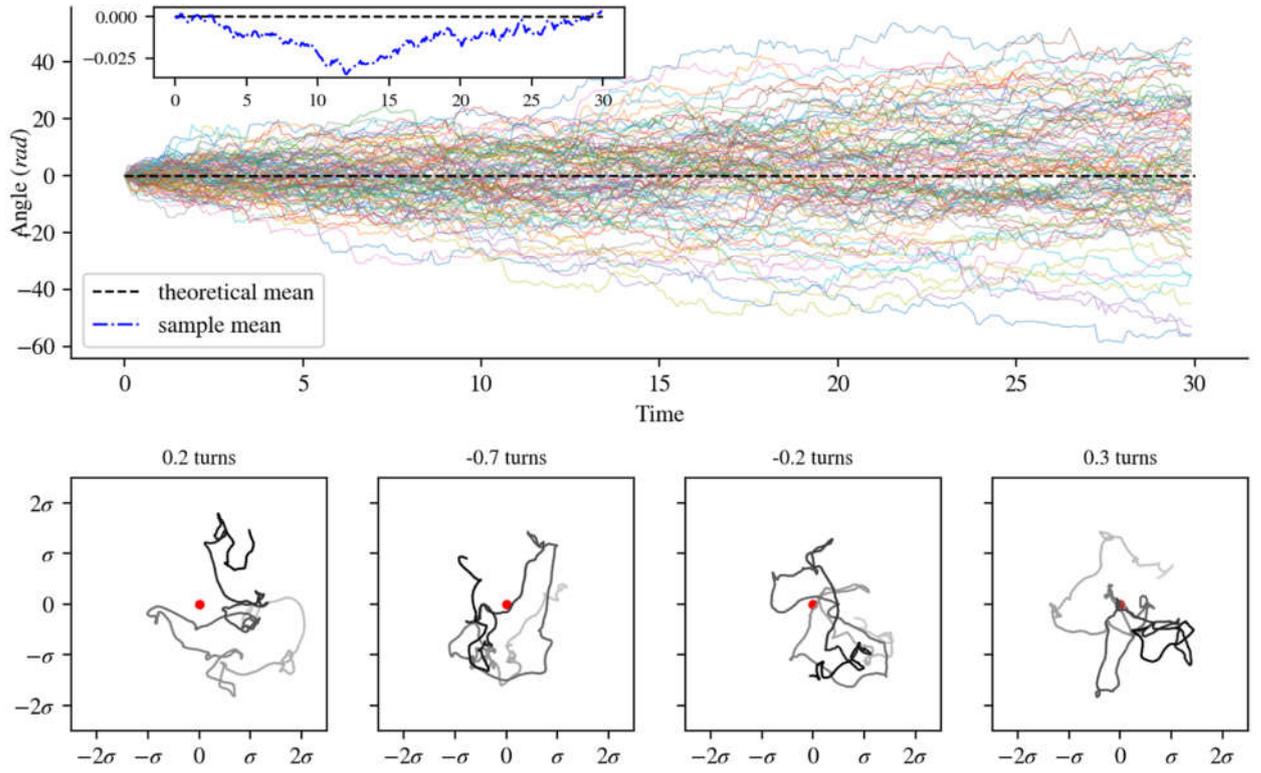

**Figure 4.** (No color in print) The same as in Figs. 1 and 2, but for parameters: $\kappa = 1000$, $\omega_0 = 20$: memoryless friction and loosely confining potential. With these parameters we are in the regime of the ordinary overdamped (no rotation) Brownian motion. The average angular speed induced by the magnetic field is small: $\langle \Omega \rangle = -3 \cdot 10^{-5}$; cf. (3.6).

The relaxation times and rotational frequencies are (resp.) $(u'^{-1}_1, u'^{-1}_2, u'^{-1}_3) = (0.0011, 0.00922, 0.2407)$ and $(u''^{-1}_1, u''^{-1}_2, u''^{-1}_3) = (-6.883, 0.840, -22.27)$. The relaxation times are much shorter than the rotation periods.